%% file: main.tex
\documentclass[sigconf,nonacm]{acmart}

\usepackage{booktabs}
\usepackage{graphicx}
\usepackage{balance}
\usepackage{datetime}
\usepackage{booktabs}
\usepackage{threeparttable}
\usepackage{multirow}
\usepackage{balance}
\usepackage{comment}
\usepackage[noend,linesnumbered,algoruled]{algorithm2e}
\usepackage{array}
\newcolumntype{?}{!{\vrule width 1.25pt}}
\usepackage{makecell}
\usepackage{varwidth}
\usepackage{url}
\usepackage{hyperref}
\usepackage{listings}
\usepackage{subfig}  
\usepackage{cleveref}
\usepackage{csquotes}
\usepackage{verbatimbox}
\usepackage{booktabs}
\usepackage{epstopdf}
\usepackage{gensymb}
\usepackage{mathtools}
\usepackage{enumerate}
\usepackage[shortlabels]{enumitem}
\usepackage{mwe}
\usepackage[normalem]{ulem}

\usepackage{epsfig}

\usepackage{caption}
\usepackage{multirow}
\usepackage{url}

\usepackage{color, colortbl}
\definecolor{LightCyan}{rgb}{0.88,1,1}
\definecolor{MistyRose}{rgb}{1,0.89,0.88}
\definecolor{PeachPuff}{rgb}{1,0.85,0.72}
\definecolor{LemonChiffon}{rgb}{1,0.98,0.80}

\usepackage{float}
\usepackage{etoolbox}
\usepackage{amsmath}

\newcommand{\tbf}{\textbf{\textcolor{red}{X}}\xspace}
\newcommand{\ignore}[1]{}

\ignore{
\usepackage{algpseudocode}
\makeatletter
\AfterEndEnvironment{algorithm}{\let\@algcomment\relax}
\AtEndEnvironment{algorithm}{\kern2pt\hrule\relax\vskip2pt\@algcomment}
\let\@algcomment\relax
\newcommand\algcomment[1]{\def\@algcomment{\footnotesize#1}}

\renewcommand\fs@ruled{\def\@fs@cfont{\bfseries}\let\@fs@capt\floatc@ruled
  \def\@fs@pre{\hrule height.8pt depth0pt \kern2pt}%
  \def\@fs@post{}%
  \def\@fs@mid{\kern2pt\hrule\kern2pt}%
  \let\@fs@iftopcapt\iftrue}
\makeatother}
\usepackage{mathtools}

\usepackage{hyperref}
\hypersetup{
  colorlinks   = true, 
  urlcolor     = blue, 
  linkcolor    = blue, 
  citecolor   = blue 
}

\newcommand{\alg}{\nit{DIVA}\xspace}

\newcommand{\mc}[1]{\mathcal{ #1}}

\newcommand{\nit}[1]{{\it #1}}
\newcommand{\val}[1]{{\sf #1}}
\newcommand{\att}[1]{{#1}}
\newcommand{\tp}[1]{{#1}}
\newcommand{\sch}[1]{\mc{#1}}

\newcommand{\base}{\nit{base}}

\newcommand{\eat}[1]{}
\newcommand{\vgh}{{\sf VGH}\xspace}
\newcommand{\vghs}{{\sf VGHs}\xspace}
\newcommand{\dgh}{{\sf DGH}\xspace}

\newcommand{\kw}[1]{#1}
\newcommand{\pair}{\kw{PAIR}-\kw{ENUM}\xspace}
\newcommand{\kas}{\kw{k}-\kw{ASSEMBLE}\xspace}

\newcommand{\lca}{\nit{lca}}
\newcommand{\gpenalty}{\nit{penalty}}

\newcommand{\attr}[1]{{\sf #1}}

\newcommand{\boxtheorem}{\hfill $\square$}

\newcommand{\red}[1]{\textcolor{red}{#1}}
\newcommand{\blue}[1]{\textcolor{blue}{#1}}
\newcommand{\purple}[1]{\textcolor{purple}{#1}}
\newcommand{\mostafa}[1]{\noindent \blue{Comment(Mostafa): #1}}
\newcommand{\fei}[1]{\noindent \red{Comment(Fei): #1}}

\newcommand{\colored}[1]{}

\newcommand{\experiment}[1]{\vspace{1mm}\noindent {\bf #1}}
\newcommand{\starter}[1]{\vspace{1mm}\noindent {\bf #1}}
\newcommand{\dataset}[1]{\vspace{1mm}\noindent \uline{#1}}
\newcommand{\stratOne}{DIVA-MinChoice\xspace}
\newcommand{\stratTwo}{DIVA-MaxFanOut\xspace}
\newcommand{\stratNaive}{DIVA-Naive\xspace}

\definecolor{cgreen}{rgb}{0.25,0.50,0.0}
\definecolor{cred}{rgb}{0.60,0.10,0.0}
\definecolor{cgray}{rgb}{0,0,0}
\definecolor{cblue}{rgb}{0.5,0.7,0.94}

\newcommand{\eattr}[1]{} 

\setcopyright{rightsretained}

\usepackage{booktabs} 

\AtBeginDocument{%
  \providecommand\BibTeX{{%
    \normalfont B\kern-0.5em{\scshape i\kern-0.25em b}\kern-0.8em\TeX}}}

\setcopyright{acmcopyright}
\copyrightyear{2018}
\acmYear{2018}
\acmDOI{10.1145/1122445.1122456}

\acmConference[CIKM '20]{29TH ACM Conference on Information and Knowledge Management}{October 19--23, 2020}{Galway, Ireland}
\acmBooktitle{WACM Conference on Information and Knowledge Management,
  October 19--23, 2020, Galway, Ireland}
\acmPrice{15.00}
\acmISBN{978-1-4503-XXXX-X/18/06}

\begin{document}
\title{Diversifying Anonymized Data with Diversity Constraints}

\author{Mostafa Milani}
\affiliation{%
  \institution{Western University}
  \city{London, Ontario, Canada}
}
\email{mostafa.milani@uwo.ca}

\author{Yu Huang}
\affiliation{%
   \institution{McMaster University}
   \city{Hamilton, Ontario, Canada}
}
\email{huang223@mcmaster.ca}

\author{Fei Chiang}
\affiliation{%
   \institution{McMaster University}
      \city{Hamilton, Ontario, Canada}
}
\email{fchiang@mcmaster.ca}

\renewcommand{\shortauthors}{}

\begin{abstract} Recently introduced privacy legislation has aimed to restrict and control the amount of personal data published by companies and shared to third parties.  Much of this real data is not only sensitive requiring anonymization, but also contains characteristic details from a variety of individuals. This diversity is desirable in many applications ranging from Web search to drug and product development.\ignore{Fairness is also a desirable property and the main component of responsible data science, which is related to diversity: fairness and impartiality for members of minority groups usually leads to diversity.} Unfortunately, data anonymization techniques have largely ignored diversity \ignore{and fairness}in its published result. This inadvertently propagates underlying bias in subsequent data analysis. We study the problem of finding a diverse anonymized data instance where diversity is measured via a set of diversity constraints. We formalize diversity constraints and study their foundations such as implication and satisfiability. We show that determining the existence of a diverse, anonymized instance can be done in PTIME, and we present a clustering-based algorithm. We conduct extensive experiments using real and synthetic data showing the effectiveness of our techniques, and improvement over existing baselines. Our work aligns with recent trends towards responsible data science by coupling \ignore{fairness and }diversity with privacy-preserving data publishing. 
\end{abstract}

\ignore{

}


\maketitle

\input{introduction}
\input{preliminaries}
\input{foundations}
\input{algorithm}
\input{experiments}

\input{related-work}
\input{conclusion}

\bibliographystyle{ACM-Reference-Format}
\bibliography{references}




\end{document}

%% file: introduction.tex
\section{Introduction} \label{sec:intro}

Organizations often share user information with third parties to analyze collective user behaviour and for  targeted marketing.  For example, in the pharmaceutical industry, hospital and medical records are shared and sold to data brokers who aggregate longitudinal data from patient records, insurance claims and lab tests to derive collective insights for research and drug development.  Protecting user privacy is critical to safeguard personal and sensitive data.  The European Union General Data Protection Regulation (GDPR), and variants such as the California Consumer Protection Act (CCPA) aim to control how organizations manage user data.  For example, a major tenet in GDPR is \emph{data minimization} that states companies should collect and share only a minimal amount of personal data sufficient for their purpose.  CCPA takes this one step further requiring companies to document and track onward transfer of data to third parties.  Given the impossibility of knowing how a published data instance will be used in the future, determining a minimal amount of personal data to share is a challenge. 

One solution is to apply differential privacy techniques to the entire data instance that provide provable guarantees.  These guarantees often rely on aggregation queries over sufficiently large samples such that the output is not influenced by the presence (or absence) of any single record \cite{dwork}.  Unfortunately, applications often experience poor data utility and accuracy due to the necessary data randomization in differential privacy.  Privacy-preserving data publishing (PPDP) provides a middle-ground to safeguard individual privacy while ensuring the published data remains practically useful for subsequent analysis.  One of the benefits of PPDP is the focus on publishing actual data, rather than statistical summaries and relationships about the data.  Anonymization is the most common form of PPDP, where quasi-identifiers and/or sensitive values are obfuscated via suppression or generalization \cite{bng-fung}.  

As anonymized instances are shared with third parties for decision making and analysis, there is growing interest to ensure that data (and the algorithms that generate and use the data) are diverse and fair.  Diversity is a rather established notion in data analytics that refers to the property of a selected set of individuals. Diversity requires the selected set to have a minimum representation from each group of individuals~\cite{drosou2017diversity,stoyanovich2018online} while determining the minimum bound for each group is often domain and user dependent. \ignore{Unlike diversity, fairness has many different interpretations in data analytics~\cite{fairnessSurvey}. The most common interpretation is {\em algorithmic fairness}~\cite{DworkFairness}. Considering a binary classifier (e.g. for admitting students at university), the goal of algorithmic fairness is to prevent discrimination against individuals in the classification task, based on their membership in some group, while maintaining utility for the classifier. A sufficient condition for algorithmic fairness is statistical parity, the property
that the demographics of those receiving positive (or negative) classifications are identical to the
demographics of the population as a whole~\cite{DworkFairness}.}

\begin{table*}[h]
\begin{minipage}[t]{0.32\textwidth}
\normalsize
\setlength\tabcolsep{2 pt}
\hspace{-2mm}\resizebox{5.8cm}{!}{
\begin{tabular}{ | l | l | l | l | l | l || l |}
\hline
 \textbf{ID}  &  \textbf{GEN} &  \textbf{ETH} &  \textbf{AGE} & \textbf{PRV} &  \textbf{CTY}  &  \textbf{DIAG}   \\
\hline \hline
$t_{1}$  & Female & Caucasian &	80 & AB & Calgary & Hypertension \\$t_2$  & Female & Caucasian &	32 & AB & Calgary & Tuberculosis \\$t_3$  & Male & Caucasian &	59 & AB & Calgary & Osteoarthritis \\ 
$t_{4}$  & Male & Caucasian &	46 & MB & Winnipeg & Migraine \\
$t_5$  & Male & African &	31 & MB & Winnipeg & Hypertension \\
$t_{6}$  & Male & African &	43 & BC & Vancouver & Seizure\\
$t_7$  & Male & Caucasian &	29 & BC & Vancouver & Hypertension \\
$t_{8}$  & Female & Asian &	58 & BC & Vancouver & Seizure \\ 
$t_9$  & Female & Asian &	47 & MB & Winnipeg & Influenza \\
$t_{10}$  & Female & Asian &	71 & BC & Vancouver & Migraine \\ 
\hline
\end{tabular}
}
\caption{\small Medical records relation ($R$)}
\label{tab:r}
\end{minipage}%
\begin{minipage}[t]{0.32\textwidth}
\normalsize
\setlength\tabcolsep{2 pt}
\resizebox{5.63cm}{!}{
\begin{tabular}{ | l | l | l | l | l | l || l |}
\hline
 \textbf{ID}  &  \textbf{GEN} &  \textbf{ETH} &  \textbf{AGE} & \textbf{PRV} &  \textbf{CTY}  &  \textbf{DIAG}   \\
\hline \hline
\colored{\rowcolor[HTML]{ffbbbb}} $r_{1}$  & $\star$ & Caucasian &	$\star$ & AB & Calgary & Hypertension \\
\colored{\rowcolor[HTML]{ffbbbb}}$r_2$  & $\star$ & Caucasian &	$\star$ & AB & Calgary & Tuberculosis \\
\colored{\rowcolor[HTML]{ffbbbb}}$r_3$  & $\star$ & Caucasian &	$\star$ & AB & Calgary & Osteoarthritis \\\hline
$r_{4}$  & Male & $\star$ &	$\star$ & $\star$ & $\star$ & Migraine \\
$r_5$  & Male & $\star$ &	$\star$ & $\star$ & $\star$ & Hypertension \\
$r_{6}$  & Male & $\star$ & $\star$ & $\star$ & $\star$ & Seizure\\
$r_7$  & Male & $\star$ &	$\star$ & $\star$ & $\star$ & Hypertension \\\hline
\colored{\rowcolor[HTML]{b3b3ff}} $r_{8}$  & Female & Asian &	$\star$ & $\star$ & $\star$ & Seizure \\
\colored{\rowcolor[HTML]{b3b3ff}} $r_{9}$  & Female & Asian &	$\star$ & $\star$ & $\star$ & Influenza \\
\colored{\rowcolor[HTML]{b3b3ff}} $r_{10}$  & Female & Asian &	$\star$ & $\star$ & $\star$ & Migraine \\
\hline
\end{tabular}
}
\caption{\small Anonymized relation with $k=3$}
\label{tab:rprime}
\end{minipage}%
\begin{minipage}[t]{0.32\textwidth}
\normalsize
\setlength\tabcolsep{2 pt}
\resizebox{5.92cm}{!}{
\begin{tabular}{ | l | l | l | l | l | l || l |}
\hline
 \textbf{ID}  &  \textbf{GEN} &  \textbf{ETH} &  \textbf{AGE} & \textbf{PRV} &  \textbf{CTY}  &  \textbf{DIAG}   \\
\hline \hline
\colored{\rowcolor[HTML]{b3b3ff}} $g_{1}$  & Female & Caucasian &	$\star$ & AB & Calgary & Hypertension \\
\colored{\rowcolor[HTML]{b3b3ff}} $g_2$  & Female & Caucasian &	$\star$ & AB & Calgary & Tuberculosis \\\hline
\colored{\rowcolor[HTML]{ffbbbb}} $g_3$  & Male & Caucasian &	$\star$ & $\star$ & $\star$ & Osteoarthritis \\ 
\colored{\rowcolor[HTML]{ffbbbb}} $g_{4}$  & Male & Caucasian &	$\star$ & $\star$ & $\star$ & Migraine \\\hline
\colored{\rowcolor[HTML]{8080fc}} $g_5$  & Male & African &	$\star$ & $\star$ & $\star$ & Hypertension \\
\colored{\rowcolor[HTML]{8080fc}} $g_{6}$  & Male & African &	$\star$ & $\star$ & $\star$ & Seizure\\\hline
$g_7$  & $\star$ & $\star$ &	$\star$ & BC & Vancouver & Hypertension \\
$g_{8}$  & $\star$ & $\star$ &	$\star$ & BC & Vancouver & Seizure \\ \hline
\colored{\rowcolor[HTML]{fc7070}} $g_9$  & Female & Asian &	$\star$ & $\star$ & $\star$ & Influenza \\
\colored{\rowcolor[HTML]{fc7070}} $g_{10}$  & Female & Asian &	$\star$ & $\star$ & $\star$ & Migraine \\ 
\hline
\end{tabular}
}
\caption{\small Anonymized relation with $k=2$.\ignore{\fei{To save space, can we put Table 1,2,4 across in a row and each of smaller size?}}}
\label{tab:r2}
\end{minipage}
\vspace{-2mm}
\caption{(a) A private relation, and its two $k$-anonymized relations in (b) and (c).}\label{tab:1}
\vspace{-8mm}
\end{table*}

To avoid biased decision making, incorporating diversity into computational models is essential to prevent and minimize discrimination against disadvantaged and minority groups\eat{~\cite{dangIR,clarkeIR,carbonellIR,agrawalIR,capanniniIR}}.  In this paper, we focus on diversity, and study how diversity requirements can be modeled and satisfied in PPDP.  \eat{during a decision making process. They are important not only for ethical and legal reasons but also for practical reasons. For example, returning diverse query answers resolves ambiguity in information retrieval tasks, and \fei{need more clear statement of how it resolves ambiguity for users}}  \eat{Diversity is also a common and effective solution to the problem of  diverse training data samples prevent over-fitting in recommendation systems~\cite{kaminskasRS,yuRS}.}  In PPDP, non-diverse data instances that exclude minority group give an inaccurate representation of the population in subsequent data analysis.  Unfortunately, early PPDP work~\cite{\ignore{AggarwalTables,meyerson,}bng-fung,SWEENEYL.2002a,SAMARATIP.2001}, and recent work on PPDP for linked data and graphs~\cite{GK16,hay} have not studied techniques to include diversity in published data instances.  Consider the following example demonstrating the challenges of applying diversity in PPDP.

\begin{example}\label{ex:intro}
Table~\ref{tab:r} shows relation $R$ containing patients medical records describing gender (\att{GEN}), ethnicity (\att{ETH}), age (\att{AGE}), province (\att{PRV}), city (\att{CTY}), and diagnosed disease (\att{DIAG}).  Third-parties such as pharmaceuticals, insurance firms are interested in an anonymized $R$ containing patients from diverse geographies, gender, and ethnicities.  Let \att{GEN}, \att{ETH}, \att{AGE}, \att{CTY}, \att{PRV}, be quasi-identifier (QI) attributes, and let \att{DIAG} be a sensitive attribute.   Existing PPDP methods such as $k$-anonymity \eat{(defined later)} prevent re-identification of an individual along the QI attributes from $k-1$ other tuples.  Table~\ref{tab:rprime} shows a $k$-anonymized instance for $k=3$ where tuples are clustered along the QI attributes via value suppression~\cite{SWEENEYL.2002a,SAMARATIP.2001}. 

The $k$-anonymization problem is to generate a $k$-anonymous relation through an anonymization process, such as generalization and suppression, while incurring minimum information loss. Suppression replaces some QI attribute values with $\star$s to achieve $k$-anonymity. There are several measures of information loss in PPDP~\cite{bng-fung}, e.g., \eat{The information loss caused by value suppression is usually measured} counting the number of $\star$s. Existing $k$-anonymization techniques do not preserve diversity in $R$ since these information loss measures do not  capture diversity semantics.  \eat{the data, e.g. ethnicity or gender diversity in $R$, since  measures of information loss can not capture diversity.}\boxtheorem
\end{example}

Unfortunately, existing methods fail to provide any diversity guarantees in published, privatized data instances.  This leads to inaccurate and biased decision making in downstream data analysis. \eat{across all sectors including public policy, education, and health care.} For example, in health care, anonymized patient records that exclude minority groups or fail to preserve the ratios of patients across different diseases misrepresent the true patient population, causing insufficient resource allocations.

To model diversity, existing work have proposed declarative methods in the form of \emph{diversity constraints}~\cite{stoyanovich2018online}, which define the expected frequencies that sensitive values in the data must satisfy.  Using $k$-anonymity as our privacy definition, and given a relation $R$, constant $k$, and a set of diversity constraints $\Sigma$, we study the problem of publishing a $k$-anonymized and diverse instance $R^*$.  An example of a diversity constraint $\sigma_1=(\att{ETH}[\nit{Asian}], 2, 5)$ requires an anonymized instance to contain a minimum of two Asian individuals and no more than five, which is satisfied by Table~\ref{tab:r} and in Table~\ref{tab:rprime}. Diversity constraints provide a declarative definition of the minimum and maximum frequency bounds that specific attribute domain values should appear in $R^*$~\cite{stoyanovich2018online}.\eat{A  fairness constraint $\phi$ requires the ratio of domain values in $R$ to be preserved in $R^*$. \eat{female to male in $R^*$ to be equal to this ratio in $R$.} 
In our example, $R_2$ contains an equal proportion of females to males, which is close to the 45-55\% ratio in $R$.  Closeness can be measured via distance function that satisfies a given threshold.  }

In this paper, we define the $(k,\Sigma)$-anonymization problem, which seeks an optimal $k$-anonymous instance $R^*$ that satisfies a set of diversity constraints, such as $\sigma_1 \in \Sigma$.  We study the $(k,\Sigma)$-anonymization decision problem, that is, whether there exists a $k$-anonymous instance $R^*$ that satisfies $\Sigma$. In Example~\ref{ex:intro}, there is no $3$-anonymized version of $R$ that satisfies  $\sigma_2=(\att{ETH}[\nit{African}],$ \ $1, 3)$ because there are only two African patients in $R$. \ignore{\eat{and they can not be preserved with $k=3$.} Similarly, there is no $3$-anonymized instance  satisfying the diversity constraints $\sigma_1$ and $\sigma_3=(\att{CITY}$ \ $[\nit{Vancouver}], 2, 4)$ tuples $t_8,t_{10}$ cannot be preserved to satisfy the both constraints. \fei{Vancouver example: If the point is about not enough tuples (same as African prev. ex), I would remove for space savings.} }


We study the validation, implication, and satisfiability problems of diversity constraints over a relation $R$, independent of PPDP, and then discuss the inherent challenges when extending to PPDP. We show that $(k,\Sigma)$-anonymization is NP-hard but the decision problem is in PTIME. \eat{prove that the approximation ratios for $(k,\Sigma)$-anonymization hold for $(k,\Sigma)$-anonymization.} We propose the \alg \ algorithm to compute a \uline{DIV}erse and \uline{A}nonymized $R^*$. \alg \ integrates anonymization with diversity by applying value suppression to find a $k$-anonymous instance satisfying a set of diversity constraints.

\starter{Contributions.} We make the following contributions: 
\begin{enumerate}[nolistsep, leftmargin=*]
    \item We study the foundations of diversity constraints; their validation, implication, satisfiability, and finding a minimal cover.  We also give an axiomatization of diversity constraints, and  present an algorithm for checking implication using this axiomatization.
    \item We define the $(k,\Sigma)$-anonymization problem that seeks a $k$-anonymous relation with value suppression that satisfies $\Sigma$.  We introduce \alg, a clustering-based algorithm that solves the $(k,\Sigma)$-anonymization problem with minimal suppression.
    \item We present two selection strategies to improve the \alg algorithm performance by selectively ordering candidate constraints and clusterings to minimize conflict and save computation. 
    \item We conduct an extensive evaluation using real data collections demonstrating the effectiveness and efficiency of our selection strategies over the naive version of \alg\, and show the utility of diversity constraints over an existing baseline. 
\end{enumerate} 

\starter{Paper Organization.} In Section~\ref{sec:prelim}, we present necessary definitions and notation.  We study foundations of diversity constraints in Section~\ref{sec:foundation}, and introduce the \alg algorithm and our selection strategies in Section~\ref{sec:alg}.  We present our evaluation results in Section~\ref{sec:exp}, related work in Section~\ref{sec:rw}, and conclude in Section~\ref{sec:conclusion}.

%% file: preliminaries.tex
\section{Preliminaries}\label{sec:prelim}


\subsection{Relations and Dependencies}
A {\em relation} $R$ with a {\em schema} $\sch{R}=\{A_1,...,A_n\}$ is a finite set of $n$-ary tuples $\{\tp{t}_1,...,\tp{t}_N\}$. \ignore{A {\em database} ({\em instance}) $D$ is a finite set of relations $R_1,...,R_m$ with schema $\mc{S}=\{\sch{R}_1,...\sch{R}_m\}$.}We denote by small letters $x,y,z$ as variables. Let $A,B,C$ refer to single attributes and $X,Y,Z$ as sets of attributes. A {\em cell} $c=t[A_i]$ is the $i$-th position in tuple $\tp{t}$ with value denoted by $c.{\sf value}$. We use $c$ to refer to $c.{\sf value}$ if it is clear from the context. Table~\ref{tab:symbols} summarizes our symbols and notations.

\begin{table}
  \centering
  \caption{Summary of notation and symbols.}\vspace{-4mm}
  \label{tab:symbols}
  \setlength\tabcolsep{2 pt}
  \small \begin{tabular}{l l}
  \toprule
    Symbol              & Description \\
    \midrule
    $R,\mc{R}$ & relation and relational schema\\
    $A,B$ & relational attributes\\
    $X,Y,Z$ & sets of relational attributes\\
    $\sqsubseteq,\star$ & suppression relation, symbol for a suppressed value\\
    $\phi,\sigma, \Sigma$ & single and set of diversity constraints \\ 
    $C,\mc{S}$ & cluster and clustering (set of clusters)\\
    \bottomrule
    \end{tabular}
\end{table}

\subsection{Privacy-Preserving Data Publishing}

$k$-anonymity\eat{is a well-known anonymization model that} prevents re-identification of an individual in an anonymized data set~\cite{SWEENEYL.2002a,SAMARATIP.2001}. Attributes in a relation are either {\em identifiers} such as SSN that uniquely identify an individual, \emph{quasi-identifier} (QI) attributes such as ethnicity, address, age that together can identify an individual, or \emph{sensitive} attributes that contain personal information.

\begin{definition}[QI-group and $k$-anonymity] \label{df:k} A relation $R$ is $k$-anonymous if every QI-group has at least $k$ tuples. A QI-group is a set of tuples with the same values in the QI attributes.\end{definition}

For example, Table~\ref{tab:rprime} has three QI-groups, $\{r_1,r_2,r_3\}$, $\{r_4,r_5,r_6,r_7\}$, and $\{r_8,r_9,r_{10}\}$, and  is $3$-anonymous.  Recent extensions of $k$-anonymity include \eat{is a prominent and the first privacy model in PPDP, there are more advanced models such as} $l$-diversity, $t$-closeness, and $(X,Y)$-anonymity, which provide improved privacy confidence (cf.~\cite{bng-fung} for a survey).   We apply  $k$-anonymity for its ease of presentation, however, our definitions and techniques can be extended to include recent PPDP models. 

\ignore{\begin{figure}
 \begin{center}
\includegraphics[width=\linewidth]{./fig/med.jpg}
\vspace{-4mm}
  \caption{\dgh and \vgh for medications.}\label{fig:med}
\vspace{-4mm}
\end{center}
\end{figure}}

\subsection{Suppression} \label{sec:suppression}

Suppression \eat{is an anonymization process that}generates an anonymized relation $R'$ from a relation $R$ by replacing some QI values in $R$ with $\star$. \eat{i.e. a special character that does not appear in $R$.} We denote this by $R \sqsubseteq R'$. Suppression clearly causes information loss which is typically measured by the number of $\star$s in $R'$.

\begin{definition}[$k$-anonymization problem~\cite{SWEENEYL.2002a}]
Given $R$, the problem of $k$-anonymization is finding $R^*$ such that (1) $R \sqsubseteq R^*$; (2) $R^*$ is $k$-anonymous; and (3) $R^*$ incurs minimum information loss. \eat{(the number of $\star$s).}\end{definition}

The $k$-anonymization problem is NP-hard for $k\ge 3$ even when QI attributes have only two values\ignore{~\cite{meyerson}} but it is in PTIME for $k=2$\ignore{~\cite{blocki}}. The best approximation for a general value of $k$ is a $O(\log k)$, and for the special case $k=3$, there is a $2$-approximation algorithm~\cite{bng-fung}.



\subsection{Diversity Constraints} \label{sec:set}
Diversity constraints are originally proposed for the set selection problem defined as follows~\cite{stoyanovich2018online}.  Given a set of $N$ items, each associated with a sensitive attribute and a utility score, the {\em set selection problem} is to select $M$ items to maximize a utility score subject to diversity constraints. The utility score is the sum of scores of each selected item. Let there be $d$ distinct values of the sensitive attribute and $m_i$ with $i \in [1, d]$ be the number of selected items with each distinct value such that $m_i \in [0,M]$ and $\sum_i(m_i)=M$. A diversity constraint $\phi$ of the  form $\nit{floor}_i \le m_i \le \nit{ceiling}_i$ specifies upper and lower bounds on $m_i$, i.e. the number of items with the $i$-th sensitive value.  These constraints ensure representation from each category known as coverage-based diversity.  To avoid tokenism, where there is only a single representative from each category, we can increase the lower bound, e.g., $m_i > 1$. Given a set of diversity constraints $\Sigma$ of the form $\phi \in \Sigma$, we define our initial problem statement.


\begin{definition}[Problem Statement ($(k,\Sigma)$-anonymization)]\label{df:k-sigma}
Consider a relation $R$ of schema $\mc{R}$, a constant $k$, a set of diversity constraints $\Sigma$. The \emph{$(k,\Sigma)$-anonymization problem} is to find a relation $R^*$ where: (1) $R \sqsubseteq R^*$, (2) $R^*$ is $k$-anonymous, (3) $R^* \models \Sigma$, and (4) $R^*$ has minimal information loss, i.e., a minimum number of $\star$'s. 
\end{definition}

%% file: foundations.tex
\section{Foundations} \label{sec:foundation}
We apply the concept of diversity constraints as proposed by Stoyanovich et. al~\cite{stoyanovich2018online} (Section~\ref{sec:set}).  We introduce a formal definition of these diversity constraints, study their validation, implication and satisfaction, define minimal cover, and present an axiomtization.  


\begin{definition} [Diversity Constraints] \label{df:div} A diversity constraint over a relation schema $\mc{R}$ is of the form $\sigma=(A[a], \lambda_l, \lambda_r)$ in which $A \in \mc{R}$, $a \in \nit{dom}(A)$ and $\lambda_l,\lambda_r$ are non-negative integers. The diversity constraint $\sigma$ is satisfied by a relation $R$ of schema $\mc{R}$ denoted $R \models \sigma$ if and only if there are at least $\lambda_l$ and at most $\lambda_r$ occurrences of the value $a$ in attribute $A$ of relation $R$. We call $[\lambda_l,\lambda_r]$ the frequency range and $A[a]$ the target value of $\sigma$. A set of diversity constraints $\Sigma$ is satisfied by $R$, denoted by $R\models \Sigma$, iff $R$ satisfies every $\sigma \in \Sigma$. \boxtheorem\end{definition}

\ignore{Following Definition~\ref{df:div}, \ignore{$(A[a], 0, \lambda_r)$ and $(A[a], \lambda_l, +\infty)$ are subsequently left-unbounded and right-unbounded diversity constraints with frequency ranges $[0,\lambda_r]$ and $[\lambda_l, +\infty)$.} we use $(A[a], \emptyset)$ to refer to a  diversity constraint with the empty frequency range. We call it a {\em false diversity constraint} since it is falsified by any relation with schema $\mc{R}$. Similarly, $(A[a], 0, +\infty)$ is a {\em true} diversity constraint that always holds.}

\subsection{Validation}\label{sec:validation}
The {\em validation problem} is to decide whether $R \models\sigma$. Assuming $\sigma=(A[a], \lambda_l, \lambda_r)$, we can run a query that counts the number of occurrences of the target value $a$ in attribute $A$ of $R$ and then check if this number lies in the frequency range $[\lambda_l, \lambda_r]$.
Diversity constraints can be extended to multiple attributes by replacing $A[a]$ with $X[t]$, where $X$ is a set of attributes and $t$ is a tuple with values from these attributes.  This extended diversity constraint $\sigma=(X[t], \lambda_l, \lambda_r)$ is satisfied by $R$ if there are at least $\lambda_l$ and at most $\lambda_r$ tuples in $R$ with the same attribute values in $t$. The validation problem for a multi-attribute diversity constraint is answered in a similar manner as the single attribute diversity constraint by extending the conditions to include each target attribute values, and aggregating the results via a count query.  Similar to traditional functional dependencies, validation is in PTIME since we can automatically generate SQL queries from the diversity constraints~\cite{FGJK08}.


\subsection{Implication and Axiomatization} \label{sec:implication}
We present an axiomatization for diversity constraints, \eat{that can be used to determine their implication.  We} and formally define the logical implication problem.

\begin{definition}[Logical Implication]
Given a set of diversity constraints $\Sigma$ over schema $\mc{R}$, and a diversity constraint $\sigma \not \in \Sigma$, we say $\Sigma$ {\em implies} $\sigma$, denoted by $\Sigma \models \sigma$, if and only if any relation $R \models \Sigma$, then $R \models \sigma$.  Given any finite set  $\Sigma$ and a constraint $\sigma$, the {\em implication problem} is to determine whether $\Sigma \models \sigma$.  
\boxtheorem\end{definition}

To test for logical implication $\Sigma \models \sigma$, and infer a new $\sigma$, we give a sound and complete axiomatization for diversity constraints.


\starter{Axiom 1 (Fixed Attributes):} If $\sigma=(X[t],\lambda_l,\lambda_r)$, $\sigma'=(X[t],\lambda'_l,\lambda'_r)$, $[\lambda_l,\lambda_r] \subseteq [\lambda'_l,\lambda'_r]$, then $\sigma \models \sigma'$.

For example, let $\sigma'=(\att{GEN}[\val{Female}], 1, 5)$, and $\sigma=(\att{GEN}[\val{Female}],$ \ $ 2, 4)$, which require [1,5] and [2,4] females, respectively.  The frequency range of $\sigma'$ subsumes the range of $\sigma$, indicating that $\sigma$ is more restrictive.  Thus, if a relation $R$ satisfies $\sigma$, it also satisfies $\sigma'$.  


\starter{Axiom 2 (Attribute Extension):} Let $\sigma=(X[t], \lambda_l, \lambda_r)$, $\sigma'=(X'[t'], 0, \lambda_r)$, $X[t] \subset X'[t']$, then $\sigma \models \sigma'$.

Intuitively, if we add new target attribute values to a satisfied constraint, we cannot guarantee that there exist tuples with the added values ($\lambda_l' = 0$).  In contrast, if there exist tuples that contain the new target attribute values, their frequency would be upper bounded by $\lambda_r$. For example, if $\sigma=(\att{ETH}[ \val{Female}], 1,5)$ then we can infer $\sigma'=(\{\att{GEN},$ \ $ \att{ETH}\}[\val{Female},$ \ $ \val{Caucasian}], 0, 5)$. If there are between [1,5] females in $R$, we can infer at most $5$ are possibly Caucasian, but cannot state there is at least one Caucasian (i.e., the individuals may be of different ethnicity).

\eat{The description doesn't match the axiom, unless I am mis-understanding.  The description sounds like the implication is the other way around $\sigma \models \sigma'$.  Also, I thought the inference is on the lower bound, not the upper bound?  Our defn. of diversity constraints requires lower bound greater than 1.}

\starter{Axiom 3 (Attribute Reduction):} Let $\sigma=(X[t], \lambda_l, \lambda_r)$, $\sigma'=(X'[t'], \lambda_l, +\infty)$, $X'[t']\subset X[t]$, then $\sigma \models \sigma'$.

Axiom 3 states that for a satisfied diversity constraint $\sigma$, if we remove a set of target attribute values from $X[t]$, we can infer at least $\lambda_l$ occurrences of the values $X'[t'] \subset X[t]$.  For example, if $\sigma = (\{\att{GEN}, \att{ETH}\}[\val{Female}, \val{Caucasian}], 1,5)$ holds over $R$, then we can conclude at least $1$ individual is female, i.e., $\sigma' = (\att{ETH}[\val{Female}], 1, +\infty)$ holds.  However, we cannot claim the number of females in $R$ is limited to $5$ since there may be individuals from other ethnicities in $R$.

\starter{Axiom 4 (Range Intersection):} Let $\sigma=(X[t],\lambda_l,\lambda_r)$, $\sigma'=(X[t]$\ $,\lambda'_l,\lambda'_r)$, then for any  $\sigma''=(X[t],\lambda''_l,\lambda''_r)$ where $[\lambda''_l,\lambda''_r] \subseteq [[\lambda_l,\lambda_r] $ \ $\cap [\lambda'_l,\lambda'_r]]$, it follows that $\{\sigma, \sigma'\} \models \sigma''$.

Intuitively, the set of tuples satisfying $\sigma, \sigma'$ would also satisfy a new diversity constraint $\sigma''$ that is more restrictive whose frequency range is the intersection of $[\lambda_l,\lambda_r]$ and $[\lambda'_l,\lambda'_r]$.  For example, let $\sigma=(\att{GEN}[\val{Female}], 1, 5)$ and $\sigma'=(\att{GEN}[\val{Female}], 3, 7)$, we can infer  $\sigma''=(\att{GEN}[\val{Female}], 3, 5)$.

\begin{theorem} The axiomatization (Ax. 1-4) is sound and complete.\end{theorem}

\emph{Proof Sketch.} Axioms 1-4 are sound as shown with the above examples. The axiomatization is also complete since any constraint $\sigma$ that can be inferred from $\Sigma$ can be obtained by applying Axioms 1-4 in a sequence.  We can prove this by showing for any $\sigma'$ that does not follow from $\Sigma$ via these axioms is not a logical implication of $\Sigma$, i.e., by construction of a relation that satisfies $\Sigma$ but not $\sigma'$.

We present Algorithm~\ref{alg:implies} that tests for logical implication by applying Axioms~1-4, i.e., checking whether $\Sigma \models \sigma$. The algorithm starts with a diversity constraint $(X[t], 0,+\infty)$ with the most general target range $\delta=[0,+\infty)$ (Line~\ref{ln:infty}), which is satisfied by any relation, hence, is inferred from $\Sigma$. The algorithm iterates over each constraint in $\Sigma$ to find constraints $\sigma'$ with target values in $\sigma$ to infer more restricted ranges $\delta$. Using Axioms~1-3 and in Lines~\ref{ln:intersect}-\ref{ln:left}, the algorithm subsequently finds target ranges $[\lambda'_l,\lambda'_r]$, $[0,\lambda'_r]$, and $[\lambda'_l,\infty)$, and applies Axiom~4 to restrict $\delta$. If $\delta$ is included in $[\lambda_l,\lambda_r]$ (applying Axiom~1 in Line~\ref{ln:implies} and checking if $(X[t], \delta) \models \sigma$), then $\Sigma$ implies $\sigma$. 
Algorithm~\ref{alg:implies} runs in linear time w.r.t. |$\Sigma$|, and proves the implication problem can be solved in linear time.

\begin{algorithm}[tb]
\KwOut{$\Sigma \models \sigma$.}
$\delta:=[0, +\infty)$;\label{ln:infty}\\
\ForEach{$\sigma'=(X'[t'], \lambda'_l,\lambda'_r) \in \Sigma$}{
\lIf{$X'[t'] = X[t]$}{$\delta := \delta \cap [\lambda'_l, \lambda'_r]$\label{ln:intersect}}
\lIf{$X \subset X'$ {\bf and} $t \subset t'$}{$\delta := \delta \cap [0, \lambda'_r]$\label{ln:right}}
\lIf{$X' \subset X$ {\bf and} $t' \subset t$}{$\delta := \delta \cap [\lambda'_l, +\infty)$\label{ln:left}}
}
\KwRet{$\delta \subseteq [\lambda_l,\lambda_r]$};\label{ln:implies}
\caption{\nit{Implies} ($\Sigma, \sigma=(X[t], \lambda_l, \lambda_r)$)}\label{alg:implies}
\end{algorithm}

\begin{example} \label{ex:imply} Consider the following execution of Algorithm~\ref{alg:implies} to check whether $\Sigma \models \sigma$, where $\Sigma = \{\sigma', \sigma''\}$,  $\sigma'=(\att{CTY}[\val{Calgary}], 2, 10)$, $\sigma''=(\{\att{GEN},\att{ETH},\att{CTY}\}[\val{Female},$ \ $\val{Caucasian},\val{Calgary}], 4, 7)$, and  $\sigma=(\{\att{ETH},\att{CTY}\}[\val{Caucasian},\val{Calgary}], 5, 8)$.  The range $\delta$ first reduces to $[0,10]$, and then to $[4,10]$ after considering $\sigma'$ and $\sigma''$. Line~\ref{ln:implies} returns true since $[5,8] \subseteq [4,10]$, and thus, $\Sigma$ implies $\sigma$.\boxtheorem\end{example}

\subsection{Satisfiability}\label{sec:satisfy}
{\em The satisfiability problem} is to determine whether a set of constraints $\Sigma$ is {\em satisfiable}, i.e. does there exist a relation $R$ such that $R \models \Sigma$.  We can apply Axioms 1 - 4, and test whether $\Sigma$ implies the false diversity constraint $\phi$, i.e.,  $\phi = (X[t], \lambda_l, \lambda_r)$ with empty range $[\lambda_l, \lambda_r]=\emptyset$.  Since there is no relation $R$ that satisfies $\phi$, if we infer that $\Sigma \models \phi$, then there is no $R$ that satisfies $\Sigma$, and $\Sigma$ is not satisfiable.

\begin{example} Let $\Sigma=\{\sigma',\sigma''\}$, where $\sigma'=(\{\att{ETH},\att{CTY}\}$ \\ $[\val{Caucasian},\val{Calgary}], 6, 8)$ and $\sigma''=(\att{CTY}[\val{Calgary}], 1, 5)$.  Clearly, $\Sigma$ is unsatisfiable since the target ranges are not compatible for persons from Calgary.  From Algorithm~\ref{alg:implies}, we can check that $\Sigma$ implies the false constraint $\phi = (\att{CTY}[\val{Calgary}], \emptyset)$, where $\delta = \emptyset$.  Given this, we conclude that $\Sigma$ implies $\phi$, and $\Sigma$ is not satisfiable.\boxtheorem\end{example}

\subsection{Minimal Cover}\label{sec:minimal}
To avoid redundancy, it is preferable to have a {\em minimal} set of constraints that are equivalent to $\Sigma$, i.e. a {\em minimal cover} of $\Sigma$. 

\begin{definition}\emph{(Minimal Cover)}. \label{defn:mincover}
Given two sets of diversity constraints, $\Sigma$ and $\Sigma'$, we say $\Sigma'$ {\em covers} $\Sigma$, 
if for every constraint $\sigma \in \Sigma$, $\Sigma' \models \sigma$.  A {\em minimal cover} $\Sigma'$  of $\Sigma$, is a set of diversity constraints such that $\Sigma'$ covers $\Sigma$, and there is no subset of $\Sigma'$ that covers $\Sigma$.
\end{definition}

Intuitively, a set of constraints $\Sigma'$ is minimal if every constraint $\sigma' \in \Sigma'$ is necessary.  That is, there is no constraint in $\Sigma'$ such that $\Sigma' \setminus \{\sigma'\} \models \sigma'$. In  Example~\ref{ex:imply}, the set of constraints $\Sigma=\{\sigma',\sigma''\}$ is minimal since neither $\{\sigma'\} \not\models \sigma''$ nor $\{\sigma''\} \not\models \sigma'$. However, $\Sigma \cup \{\sigma\}$ is not minimal since $\{\sigma',\sigma''\} \models \sigma$ and $\sigma$ is redundant.  We can check the minimality of a set of constraints $\Sigma$ using Algorithm~\ref{alg:implies}, by testing the logical implication of every constraint in $\Sigma$. 


In the remainder of the paper, we assume $\Sigma$ is satisfiable and minimal. We verify and reject unsatisfiable sets of constraints, and verify minimality by removing redundant constraints.

\subsection{$(k,\Sigma)$-Anonymization: Decision Problem} \label{sec:anon}

We now turn to the decision problem of $(k,\Sigma)$-anonymization, and show that the decision problem is tractable but unfortunately, the problem in Defn~\ref{df:k-sigma} is not. First, given our updated Defn.~\ref{df:div} of diversity constraints, we update $\Sigma$ in our $(k,\Sigma)$-anonymization problem statement in Defn~\ref{df:k-sigma} to reflect these constraints.  

\eat{ 
Now we use diversity constraints to represent diversity in data anonymization and data publishing. We define the $(k,\Sigma)$-anonymization problem that considers diversity constraints $\Sigma$.

\begin{definition}[$(k,\Sigma)$-anonymization] \label{df:pr} Consider a relation $R$ of schema $\mc{R}$, a constant $k$, a set of diversity constraints $\Sigma$. The $(k,\Sigma)$-anonymization problem is to find a relation $R^*$ that has the following properties: (1) $R \sqsubseteq R^*$, (2) $R^*$ is $k$-anonymous, (3) $R^* \models \Sigma$, and (4) $R^*$ has minimal information loss, i.e., minimum number of $\star$'s. 
\boxtheorem\end{definition}
}

\starter{The Decision Problem.}
Given relation $R$, value $k$, diversity constraints $\Sigma$, the $(k,\Sigma)$-\emph{anonymization decision problem} is to decide whether there exists an $R^*$ such that:  (1) $R \sqsubseteq R^*$; (2) $R^*$ is $k$-anonymous; and (3) $R^* \models \Sigma$.

We assume for any constraint $\sigma=(X[t], \lambda_l, \lambda_r)$ in $\Sigma$, $\lambda_l \ge k$. Constraint $\sigma$ can only be satisfied when the frequency of value $a$ is greater than or equal to $k$ due to the $k$-anonymity condition in $R^*$.

\begin{theorem}\label{th:ptime}
Consider relation $R$, value $k$, and a constraints $\Sigma$. The $(k,\Sigma)$-anonymization decision problem is in PTIME w.r.t. \eat{the size of}$|R|$.\boxtheorem\end{theorem}

\emph{Proof Sketch.} The proof of Theorem~\ref{th:ptime} is based on a naive algorithm that exhaustively checks every possible clustering of tuples in $R$ to generate $X$-groups that satisfy $\Sigma$. Since the number of possible clusterings is polynomial in the size of $R$, and exponential in the size of $\Sigma$, the decision problem is tractable. In Section~\ref{sec:alg}, we propose an algorithm for solving the $(k,\Sigma)$-anonymization decision problem by optimizing the naive algorithm, i.e., our new algorithm generates a $k$-anonymized instance $R^*$ that satisfies $\Sigma$.
 
\begin{proposition} \label{pr:np} Consider relation $R$, value $k$, and constraints $\Sigma$. The $(k,\Sigma)$-anonymization problem is NP-hard w.r.t. $|R|$.\boxtheorem\end{proposition}

\emph{Proof Sketch.} The $(k,\Sigma)$-anonymization problem extends the $k$-anonymization problem, which is proved to be NP-hard~\cite{bng-fung}\ignore{meyerson}.

%% file: algorithm.tex
\section{The DIVA Algorithm} \label{sec:alg}

We present the \uline{DIV}ersity and \uline{A}nonymization algorithm (\alg) that solves the  $(k,\Sigma)$-anonymization problem. \alg takes as input a relation $R$, a minimal and satisfiable set of diversity constraints $\Sigma$, constant $k$, and returns a $k$-anonymous and diverse relation $R'$ that satisfies $\Sigma$. \alg is a clustering-based anonymization algorithm that works in two phases: (i) {\em clustering}, by partitioning $R$ into disjoint clusters of size $\ge k$; and (ii) {\em suppression}, by suppressing a minimal number of QI values in each cluster such that they have the same QI values, and form a QI-group of size $\ge k$. The result is a $k$-anonymous relation, as every QI-group is of size $\ge k$. 

Algorithm~\ref{alg:diva} presents the \alg algorithm details. In the clustering phase in Line~\ref{ln:d-div}, \alg uses the \nit{DiverseClustering} procedure to generate a set of diverse clusters $\mc{S}_\Sigma$. These clusters guarantee that the diversity constraints in $\Sigma$ will be satisfied by $R_\Sigma$ after the suppression phase in Line~\ref{ln:d-suppress}. If there is no $k$-anonymous relation $R'$ that satisfies $\Sigma$, there is no such clustering, and \alg returns $\mc{S}_\Sigma:=\emptyset$. We provide details of \nit{DiverseClustering} in Section~\ref{sec:div-cluster}. 


In the suppression phase, \alg suppresses values according to the clusters in $\mc{S}_\Sigma$. The \nit{Suppress} procedure iterates over tuples in each cluster of $\mc{S}$, and suppresses $A_i$ attribute values if there is more than one value for $A_i$ in the same cluster. Assuming each cluster in $\mc{S}$ contains at least $k$ tuples, the result of \nit{Suppress} in $R$ is a $k$-anonymous relation.

Returning to Algorithm~\ref{alg:diva}, \alg anonymizes the remaining tuples of $R$ that are not in $\mc{S}_\Sigma$ (Line~\ref{ln:d-for}) by applying an existing $k$-anonymization algorithm (Line~\ref{ln:d-opt}). \alg is amenable to any $k$-anonymization algorithm. In Line~\ref{ln:d-integrate}, \nit{Integrate} returns $R'=R_\Sigma \cup R_k$ if $R'\models \Sigma$. Otherwise, $R'$ falsifies the upper bounds of some of the constraints in $\Sigma$ because of $R_k$, and \nit{Integrate} resolves this by suppressing minimal values in $R'$ to satisfy $\Sigma$.

\begin{algorithm}[tb]
\KwOut{$k$-anonymous and diverse relation.}
$\mc{S}_\Sigma:=\nit{DiverseClustering}(R, \Sigma, k)$;\label{ln:d-div}\\
\lIf{$\mc{S}_\Sigma = \emptyset$}{\KwRet{{\bf unsatisfiable}}\label{ln:d-if}}
$R_\Sigma:=\nit{Suppress}(\mc{S}_\Sigma)$;\label{ln:d-suppress}\\

\lForEach{$C_i \in \mc{S}_\Sigma$}{$R := R \setminus C_i$\label{ln:d-for}}
$R_k:=\nit{Anonymize}(R, k)$;\label{ln:d-opt}\\
\KwRet{$\nit{Integrate}(R_\Sigma, R_k)$};\label{ln:d-integrate}
\caption{\alg($R, \Sigma, k$)}\label{alg:diva}
\end{algorithm}

\eat{ We first explain the general idea of the algorithm with an example, then we describe its detail. In \alg, the clustering phase (i) has two steps:
    \item In the first step, the algorithm generates a set of diverse clusters $\mc{S}_\Sigma$ by calling \nit{DiverseClustering} in Line~\ref{ln:div}. $\mc{S}_\Sigma$ guarantees the diversity constraints $\Sigma$ will be satisfied after the suppression phase. If there is no $k$-anonymous relation $R'$ that satisfied the constraints, there is no such clustering, \nit{DiverseClustering} returns $\mc{S}_\Sigma:=\emptyset$, and \alg responds with an error message (Line~\ref{ln:if}).
    
    \item In the second step of the clustering phase, \alg generates an optimal clustering $\mc{S}_\nit{OPT}$ from the tuples that are not in $\mc{S}_\Sigma$. In particular in Line~\ref{ln:for}, \alg removes the tuples of $R$ that appear in the clusters in $S_\Sigma$, then calls \nit{OptimalClustering} to find $\mc{S}_\nit{OPT}$. \nit{OptimalClustering} can be the clustering technique in any existing clustering-based $k$-anonyzmiation algorithm. In Section~\ref{sec:exp}, we use the clustering in the $k$-member anonymization algorithm in~\cite{byun2007efficient}.\mostafa{Yu, is there a reason you chose~\cite{byun2007efficient}? It is good if we can say a sentence about it here.}
    
}

\begin{example} \label{ex:alg} Consider relation $R$ in Table~\ref{tab:r}, $k=2$, and $\Sigma = \{\sigma_1, \sigma_2, \sigma_3\}$, where $\sigma_1=(\att{ETH}[\nit{Asian}], 2, 5)$, $\sigma_2=(\att{ETH}[\nit{African}], 1, 3)$ and $\sigma_3=(\att{CTY}[\nit{Vancouver}], 2, 4)$.  \nit{DiverseClustering} returns a clustering $\mc{S}_\Sigma=\{C_1, C_2, C_3\}$ where $C_1=\{t_9,t_{10}\}$,  $C_2=\{t_5,t_6\}$, and  $C_3=\{t_7,t_8\}$. Tuples $t_9,t_{10}$ contain the same value $\att{ETH}=\nit{Asian}$, and together with $C_1$ guarantee that the lower bound in $\sigma_1$ will be satisfied. $C_2$ and $C_3$ satisfy the lower bounds of $\sigma_2$ and $\sigma_3$ for $\att{ETH}=\nit{African}$ and $\att{CTY}=\nit{Vancouver}$, respectively.  Note that other clusterings, which satisfy $\Sigma$, are possible, such as $\{C_2, \{t_8,t_{10}\}\}$.  In  Section~\ref{sec:div-cluster}, we describe how we select one of these clusterings.



\nit{DiverseClustering} returns an empty set if there is no clustering that satisfies $\Sigma$.  For example, if $k=3$ there is no possible anonymization that satisfies $\sigma_1,\sigma_3$. In particular, there are no clusters of size $3$ that preserve both Vancouver and Asian. For $k=2$, \alg continues with the \nit{Suppress} procedure that transforms the tuples in $\mc{S}_\Sigma$ to $R_\Sigma = \{g_5,...,g_{10}\}$ as shown in Table~\ref{tab:r2}.  \alg anonymizes the remaining tuples $R \setminus \mc{S}_\Sigma = \{t_1,t_2,t_3,t_4\}$ using an existing $k$-anonymization algorithm that minimizes the number of $\star$s. In this case, the optimal result is $R_k=\{g_1,g_2,g_3,g_4\}$ in Table~\ref{tab:r2}. The \nit{Integrate} procedure returns $R_\Sigma \cup R_k$, which satisfies $\Sigma$.

\nit{Integrate} resolves any inconsistency caused by adding $R_k$. For example, if $\Sigma=\{\sigma_1,...,\sigma_4\}$ in which $\sigma_4=(\att{GEN}[\nit{Male}], 1,3)$, $R_\Sigma \cup R_k \not\models \sigma_4$ because there are 4 males in $R_\Sigma \cup R_k$. \nit{Integrate} suppresses \att{GEN} in $g_5,g_6$ or $g_3,g_4$ to satisfy $\sigma_4$.
\boxtheorem\end{example}

\subsection{Diverse Clustering} \label{sec:div-cluster}
We now describe the \nit{DiverseClustering} routine in the \alg algorithm, and define a clustering that satisfies a diversity constraint.

\begin{definition} Given a diversity constraint $\sigma$ over a relation $R$ and a clustering $\mc{S}$ with clusters of tuples in $R$, $\mc{S}$ satisfies $\sigma$, denoted as $\mc{S} \models \sigma$ if $\nit{Suppress}(\mc{S}) \models \sigma$. The clustering $\mc{S}$ satisfies a set of constraints $\Sigma$, if $\mc{S}\models \sigma_i$ for every $\sigma_i \in \Sigma$.\end{definition}

In Example~\ref{ex:alg}, $\mc{S}=\{C_1\}$ satisfies $\sigma_1$ since $\nit{Suppress}(\mc{S})=\{g_9,g_{10}\}$ (cf. Table~\ref{tab:r2}) satisfies $\sigma_1$.  The objective of \nit{DiverseClustering} is to find $\mc{S}_\Sigma$ that satisfies $\Sigma$. This works by computing clustering $\mc{S}_{\sigma_i}$ that satisfy diversity constraints $\sigma_i \in \Sigma$, and then computing $\mc{S}_\Sigma$ by merging the clusterings 
$\mc{S}_{\sigma_i}$. The main challenge is to ensure the clustering for each $\sigma_i$ is consistent with clusterings for the other constraints in $\Sigma$. If so, this allows us to merge the $\mc{S}_{\sigma_i}$ to obtain $\mc{S}_{\Sigma}$.  

\begin{definition}[Consistent clusterings]\label{df:conflict} Consider diversity constraints $\sigma_i$ and $\sigma_j$ over relation $R$. Two clusterings $\mc{S}_{\sigma_i}$ and $\mc{S}_{\sigma_j}$ are consistent if and only if $\mc{S}_{\sigma_i} \models \sigma_i$ and $\mc{S}_{\sigma_j} \models \sigma_j$ implies $\nit{Merge}(\mc{S}_{\sigma_i}, \mc{S}_{\sigma_j}) \models \{\sigma_i, \sigma_j\}$.\end{definition}

\nit{Merge} in Defn.~\ref{df:conflict} merges clusters if they overlap, otherwise their union is computed, e.g., $\nit{Merge}(\{\{t_5,t_6\}\}, \{\{t_6,t_7\}\})=\{\{t_5,t_6,t_7\}\}$, and $\nit{Merge}(\{\{t_5,t_6\}\}, \{\{t_7,t_8\}\})=\{\{t_5,t_6\}, \{t_7,t_8\}\}$.  We can check the consistency of two clusterings using \nit{Merge} and \nit{Suppress}.



\begin{example} In Example~\ref{ex:alg}, $\mc{S}_2=\{\{t_5,t_6\}\}$ and $\mc{S}_3=\{\{t_6,t_7\}\}$ are not consistent w.r.t $\sigma_2$ and $\sigma_3$, because $\mc{S}_2 \models \sigma_2$ and $\mc{S}_3 \models \sigma_3$, but $\nit{Merge}(\mc{S}_2,\mc{S}_3)=\{\{t_5,t_6,t_7\}\} \not\models \{\sigma_2, \sigma_3\}$. This occurs since $t_6$ appears in two different clusters $\{t_5,t_6\}$ and $\{t_6,t_7\}$ in $\mc{S}_2$ and $\mc{S}_3$, respectively.  Consequently, the value \nit{Vancouver} will be suppressed in the clustering $\nit{Merge}(\mc{S}_2,\mc{S}_3)$ because  $t_6[\att{CTY}]\neq t_5[\att{CTY}]$, and hence, $\sigma_3$ will not be satisfied. \boxtheorem\end{example}


It is straighforward to show that if $\mc{S}_{\sigma_i} \models \sigma_i$ for every $\sigma_i \in \Sigma$, and every pair of $\mc{S}_{\sigma_i},\mc{S}_{\sigma_j}$ are consistent, we can generate $\mc{S}_\Sigma \models \Sigma$ by merging all clusterings $\mc{S}_{\sigma_i}$. Note that it is not necessary to check consistency of every pair of clusterings $\mc{S}_{\sigma_i},\mc{S}_{\sigma_j}$, as we only need to check if $\sigma_i,\sigma_j$ apply to some tuples that are common to both constraints. We use this intuition to transform our problem of computing all $\mc{S}_{\sigma_i}$ to the problem of graph coloring.


\subsubsection{Modeling as Graph Coloring.} 
We model the problem of finding the clusterings $\mc{S}_{\sigma_i}$ as a graph coloring problem. Given an undirected graph $G = (\Gamma,E)$, where $\Gamma$ and $E$ denote the set of vertices and edges, respectively, and $m$ distinct colors, the graph coloring problem is to color all vertices subject to certain constraints. In its simplest form, no two adjacent vertices can have the same color. 

\begin{figure}
    \centering
    \includegraphics[width=5.8cm]{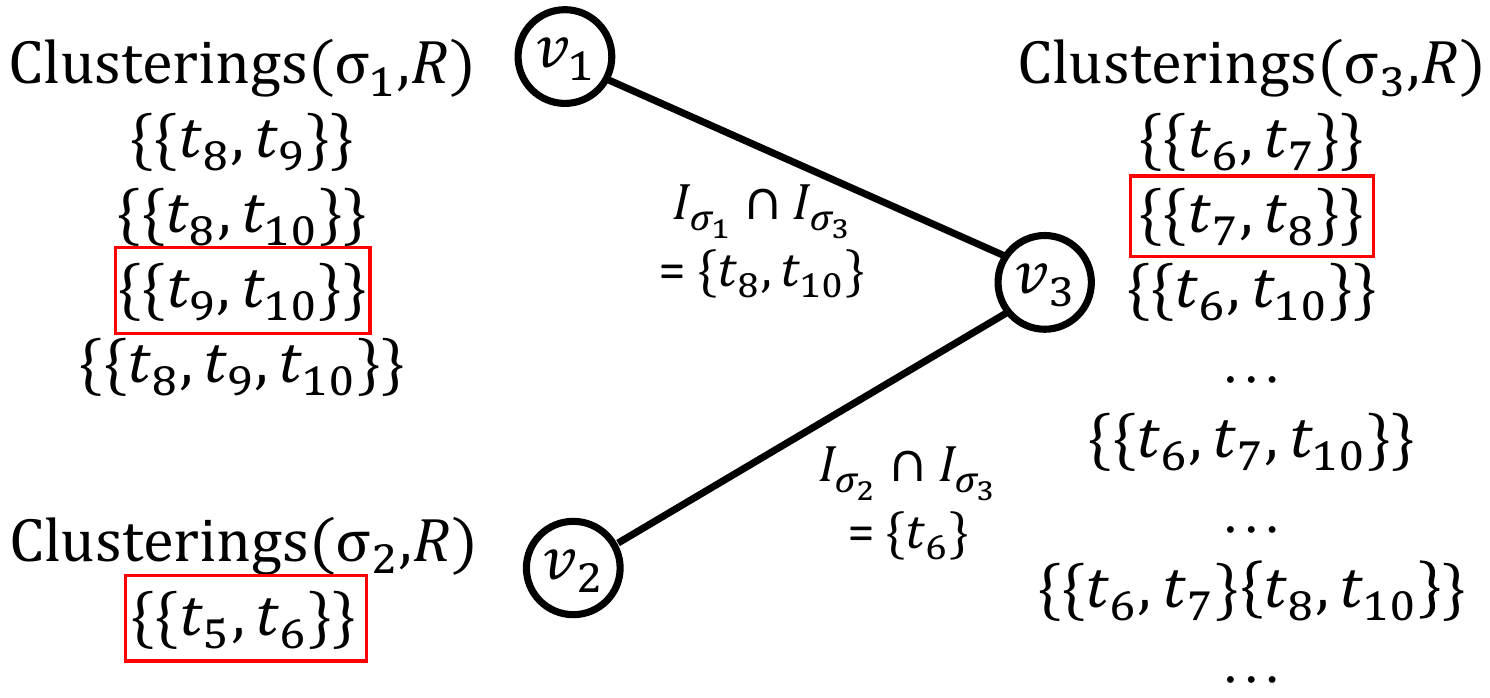}
    \vspace{-4mm}
    \caption{Diverse clustering as graph coloring.}
    \label{fig:graph}
\end{figure}

For relation $R$ and diversity constraints $\Sigma$, we model each diversity constraint $\sigma_i \in \Sigma$ as a vertex $v_i \in \Gamma$.  We use $v_i.\nit{constraint}$ to refer to $\sigma_i$. We define the  \emph{relevant tuples} of $\sigma_i$, denoted $I_{\sigma_i} \subseteq R$, as tuples containing the target values of $\sigma_i$.  We record the relevant tuples of $\sigma_i$ in vertex $v_i$.  An edge $e_{ij} \in E, e_{ij} = (v_i, v_j)$, exists between vertices $v_i$ and $v_j$ if there is at least one tuple in the intersection of their relevant tuple sets, i.e., ($I_{\sigma_i} \cap I_{\sigma_j}) \neq \emptyset$. In Example~\ref{ex:alg}, $G$ contains three vertices corresponding to $\sigma_1,\sigma_2,\sigma_3$ (cf. Figure~\ref{fig:graph}), and two edges $E = \{(v_1, v_3), (v_2, v_3\}\}$.  The relevant sets $I_{\sigma_1}$ = \{$t_8,t_9,t_{10}$\}, $I_{\sigma_3}$ = \{$t_6,t_7,t_8,t_{10}$\}, have a non-empty intersection of \{$t_8,t_{10}$\}.  Similarly, for $I_{\sigma_2} = \{t_5,t_6$\}, $I_{\sigma_2} \cap  I_{\sigma_3} = \{t_6$\}.  We note that $I_{\sigma_1} \cap  I_{\sigma_2} = \emptyset$. Choosing a color $c_i$ for vertex $v_i$ is analogous to finding a clustering $\mc{S}_{\sigma_i}$ for $\sigma_i$.  In our setting, to color two adjacent $v_i,v_j$, we must check that their clusterings $\mc{S}_{\sigma_i}$ and $\mc{S}_{\sigma_j}$ are consistent.  We define $c_i.\nit{clustering}$ to refer to clustering corresponding to color $c_i$.

\begin{algorithm}[tb]
\KwOut{Clustering $\mc{S}_\Sigma$.}
$G := \nit{BuildGraph}(R,\Sigma);$\label{ln:graph}\\
$V := \emptyset;\mc{S}_\Sigma := \emptyset;$\label{ln:init}\\
\If{$\nit{Coloring}(G,V,R)$\label{ln:exists}}{\lForEach{$\langle v_i, c_i\rangle \in V$\label{ln:merge}}{$\mc{S}_\Sigma := \nit{Merge}(\mc{S}_\Sigma, c_i.\nit{clustering})$}}

\KwRet{$\mc{S}_\Sigma$};
\caption{$\nit{DiverseClustering}(R, \Sigma, k)$}\label{alg:search}
\end{algorithm}

Algorithm~\ref{alg:search} presents the details of \nit{DiverseClustering}.  We build the graph $G$ for $\Sigma$ and $R$ (Line~\ref{ln:graph}).  We then initialize the clustering $\mc{S}_\Sigma$ and a mapping $V$ that stores the color (assigned clustering) for each vertex (Line~\ref{ln:init}), and checks if a coloring exists via  $\nit{Coloring}$.

\begin{algorithm}[tb]
\KwOut{{\bf true} if there exists a coloring of $G$, otherwise {\bf false}.}
\lIf{$V$ {\bf contains all vertices of} $G$}{\KwRet{\bf true}\label{ln:vtrue}}
$v := \nit{NextVertex}(G,V)$;\label{ln:next}\\
\ForEach{$\mc{S} \in \nit{Clusterings}(v.\nit{constraint},R)$\label{ln:candidfor}\label{ln:forclst}}{
$\nit{consistent} := {\bf true}$;\label{ln:cnf}\\
\ForEach{$\langle v',c'\rangle \in V$ {\bf s.t.} $v'$ {\bf is adjacent to} $v$}{
\If{$\mc{S}$ {\bf and} $c'.\nit{clustering}$ {\bf are inconsistent}}{$\nit{consistent} := {\bf false};{\bf break};$}}
\If{$\nit{consistent}$}{
$c := $ {\bf new color with clustering} $\mc{S};$\\
$V := V \cup \{\langle v,c\rangle\};$\label{ln:initfor}\\
\lIf{$\nit{Coloring}(G, V,R)$\label{ln:candidif}}{
\KwRet{\bf true}\label{ln:iftrue}
}
$V := V \setminus \{\langle v,c\rangle\};$\label{ln:closefor}
}

}

\KwRet{\bf false}\label{ln:retfalse}

\caption{$\nit{Coloring}(G,V,R)$}\label{alg:coloring}
\end{algorithm}


Algorithm~\ref{alg:coloring} presents the recursive function, $\nit{Coloring}$, that takes a graph $G$, the mapping $V$ (specifying the colored vertices), relation $R$, and returns {\em true} if the remaining vertices of $G$ can be colored; otherwise it returns \emph{false}. In the naive version, \nit{Coloring} randomly selects an uncolored vertex (Line~\ref{ln:next}) to color using \nit{NextVertex}. In Section~\ref{sec:strategies}, we present two strategies for selecting candidate vertices.  Given a vertex $v$, we try to color $v$  by checking whether the candidate clustering of $v$ and its adjacent vertices are inconsistent (Lines~\ref{ln:forclst}-\ref{ln:closefor}).  The routine $\nit{Clusterings}$ returns candidate clusterings $\mc{S}$ that satisfy $v.\nit{constraint}$ ($\nit{Suppress}(\mc{S}) \models v.\nit{constraint}$).  For example, in Example~\ref{ex:alg}, $\nit{Clusterings}(\sigma_1,R)$ contains four different clusterings $\{\{t_8,t_9\}\}$, $\{\{t_8,t_{10}\}\}$, $\{\{t_9,t_{10}\}\}$, $\{\{t_8,$ \ $t_9,t_{10}\}\}$, while $\nit{Clusterings}(\sigma_2,R)$ contains only one clustering $\{\{t_5,$ \ $t_6\}\}$.  In the naive algorithm, we assume $\nit{Clusterings}$ returns clusterings in random order.  We present strategies in Section~\ref{sec:strategies} to order the clusterings to minimize inconsistencies. In Lines~\ref{ln:cnf}-\ref{ln:closefor}, we check whether $\mc{S}$ has inconsistency with the clustering of any constraint modeled by an adjacent vertex $v'$ to $v$.  If they are consistent, we generate a new color $c$ assigned to the clustering $\mc{S}$, and we temporarily color $v$ with $c$ by adding $\langle v,c\rangle$ to $V$.  We then recursively call \nit{Coloring} to check whether the remaining vertices in $G$ can be colored. If the color $c$ does not work, i.e. \nit{Coloring} returns false in Line~\ref{ln:iftrue}, we remove $\langle v,c \rangle$ from $V$, and try another color. If all clusterings are inconsistent, i.e., there is no successful coloring of $v$, we return false in Line~\ref{ln:retfalse}, to backtrack and evaluate a different vertex.

\eat{The function then tries to color $v$ with different possible colors Lines~\ref{ln:forclst}-\ref{ln:closefor}. This is done by selecting a candidate clustering $\mc{S}$ that satisfies $v.\nit{constraint}$ ($\nit{Suppress}(\mc{S}) \models v.\nit{constraint}$). Here, $\nit{Clusterings}$ returns the set of all such clusterings. In Exapmle~\ref{ex:alg}, $\nit{Clusterings}(\sigma_1,R)$ contains four different clusterings $\{\{t_4,t_5\}\}$, $\{\{t_4,t_6\}\}$, $\{\{t_5,t_6\}\}$, $\{\{t_4,t_5,t_6\}\}$, and  $\nit{Clusterings}(\sigma_2,R)$ contains only one clustering $\{\{t_8,t_{10}\}\}$.} 

\begin{example} Consider an execution of Alg.~\ref{alg:coloring} \nit{Coloring} on the graph $G$ in Figure~\ref{fig:graph}, with vertices \{$v_1, v_2, v_3$\} representing constraints \{$\sigma_1, \sigma_2, \sigma_3$\}, respectively.  The candidate clusterings that satisfy each constraint (i.e., the output of the routine \nit{Clusterings}) are shown beside each vertex.  Consider vertex $v_1$ first (Line~\ref{ln:next}), and we select $S_{\sigma_1}=\{t_9,t_{10}\}$, which is consistent with any other clustering.  We then try to color vertices $v_2$ and $v_3$ by recursively calling \nit{Coloring} in Line~\ref{ln:iftrue}.  If vertex $v_2$ is selected, the only clustering is $\mc{S}_{\sigma_2}=\{\{t_5,t_{6}\}\}$ that is consistent with $\mc{S}_{\sigma_1}$.  Considering the last vertex $v_3$, we iterate over the clusterings for $\sigma_3$, and determine that the only consistent clustering (w.r.t. $\mc{S}_{\sigma_1}$ and $\mc{S}_{\sigma_2}$) is $\{\{t_7,t_8\}\}$, which we assign to $\mc{S}_{\sigma_3}$.  Since we have found a clustering satisfying all constraints (i.e., a coloring of all vertices), the \nit{Coloring} routine returns {\em true} with $V$ containing the vertices and their colors (i.e., clusterings). The calling routine \nit{DiverseClustering} uses $V$ to compute the final clustering as $\mc{S}_\Sigma=\{\{t_5,t_6\},\{t_7,t_8\},\{t_9,t_{10}\}\}$.
\boxtheorem\end{example}


\eat{In doing so, we reduce   the computation time with \blue{no} impact on accuracy. }

\starter{Runtime Analysis.} \alg runs in polynomial time w.r.t. $|R|$ since \nit{DiverseClustering}, \nit{Anonymize}, and \nit{Suppress} run in polynomial time. \eat{In particular, the $k$-member algorithm in \nit{Anonymize} is an efficient algorithm and the routines \nit{Suppress},} \nit{DiverseClustering} and its recursive procedure \nit{Coloring} run in polynomial time w.r.t. $|R|$ since the number of candidate clusterings for each constraint is polynomial w.r.t. $|R|$. In particular, the size of these clusters is in $[k,2k-1]$ and there are polynomially many clusters of each size. Note that there is no cluster of size $\ge 2k$ because we can split them into clusters of size $\ge k$. \nit{DiverseClustering} and \alg run in exponential time w.r.t. $|\Sigma|$ since we can assign $O(|R|)$ different clusterings to each constraint.  In the next section, we present strategies to improve the performance of \nit{Coloring} while evaluating the space of possible assignments.  

\subsection{Selection Strategies}
\label{sec:strategies}
In the naive version of \alg, we randomly select a constraint and a clustering to evaluate.  These choices impact algorithm performance as poor initial selections can lead to increased backtracking operations downstream.  \eat{Given apriori knowledge of constraints,}  We selectively order the constraints (vertices) and clusterings (colors) that most likely lead to a graph coloring while minimizing the need to backtrack.  We start evaluating  constraints (vertices) that are the most difficult to satisfy.  By postponing these candidates, we may encounter fewer or no possible consistent clusterings as we assign clusterings to less restrictive constraints.  We apply this intuition to propose the following two strategies.
\noindent {\bf \stratOne:} Our preference is to select constraints with the fewest  candidate clusterings, as we start with the most restrictive constraints first, i.e., those with the fewest choices, ensuring that these constraints are first satisfied.  In the routine \nit{NextVertex}, we initially select a vertex $v$ \eat{that has the smallest number of possible clusterings. At the beginning this is a vertex} with a minimum value  $|\nit{Clusterings}(v.\nit{constraint}, R)|$. As we visit vertices and assign (colors) clusterings, we update the candidate clusterings for their neighbors.  

\noindent {\bf \stratTwo:} In this strategy, we target constraints that overlap with the highest number of other constraints. This is modeled in the graph $G$ as vertices with the maximum number of unvisited edges. We preferentially select these constraints due to their high number of interactions with other constraints, which lead to an increased number of target attributes, and bounds that the relevant tuples must satisfy. This heuristic strategy aims to satisfy ``maximum overlap'' constraints first, and perform early pruning of unsatisfiable clusterings to reduce the number of clustering evaluations downstream. The vertex selection in this strategy is similar to {\em incidence degree ordering} in graph coloring~\cite{ido-coloring}.
    

In both strategies, \nit{Clusterings} returns a list of clusterings in ascending order of the number of overlapping tuples.\eat{between the clusterings and the target of the neighboring constraints.} For instance, for a clustering $\mc{S}$ and a neighboring vertex $v$ (constraint $\sigma$), overlapping tuples are in the target $I_\sigma$ and in a cluster in $\mc{S}$.  In Section~\ref{sec:exp_perf}, we show these strategies improve runtime by an average 24\%.


\begin{example} In Fig.~\ref{fig:graph}, the \stratOne strategy first selects vertex $v_2$ ($\sigma_2$),  since $|\nit{Clusterings}(\sigma_1,R)|=4$, $|\nit{Clusterings}(\sigma_2,R)|=1$, $|\nit{Clusterings}(\sigma_3,R)|=12$.  \eat{and $\sigma_1$ has the minimum number of clusterings.} After assigning cluster $\{\{t_5,t_{6}\}\}$ to $v_2$, we update the clusterings, and vertices $v_1,v_3$ will each have 4 clusterings; we break ties randomly.  In \stratTwo, we first select vertex $v_3$ ($\sigma_3$) containing two unvisited edges.  \nit{Clusterings} then computes cluster $\{\{t_7,t_8\}\}$ has 2 overlapping tuples ($t_8,t_{10}$ are in $I_{\sigma_1}$).  Similarly, cluster $\{\{t_7,t_8\}\}$ has 1 overlapping tuple $t_8$ in $I_{\sigma_1}$.  Hence, clustering $\{\{t_7,t_8\}\}$ is ranked first assuming it wins the tie against clustering $\{\{t_7,t_{10}\}\}$. \eat{which also has 1 overlapping tuple.}  We randomly select between $v_1$ and $v_2$ given their equal number of unvisited  edges.
\boxtheorem\end{example}










%% file: experiments.tex
\section{Experiments}\label{sec:exp}
Our evaluation has the following objectives:
(1) We evaluate \alg's accuracy using three types of diversity constraints as we vary $k$, and the conflict rate among tuples.  
(2) We evaluate the accuracy and performance of all \alg variants as we vary $k, |\Sigma|$, the conflict rate, and the target attribute(s) data distribution.
(3) We compare against an existing $k$-anonymization baseline algorithm to evaluate the cost of introducing diversity constraints into data anonymization. 


\subsection{Experimental Setup}
We implement \alg using Python 3.6 on a server with 32 Core Intel Xeon 2.2 GHz processor with 32GB RAM.  We describe the datasets, diversity constraints, and baseline comparative algorithm.

\begin{table}
\begin{center}
\small{
\centering
\caption{Data characteristics.}
\label{tb:dataset}
\vspace{-4mm}
\begin{tabular}{l|cccc}
\toprule
{\bf }  & {\bf Pantheon}		& {\bf Census} & {\bf Credit} &  {\bf Population (Syn)}	\\
\midrule
$|R|$		&	11,341		&	299,285	& 1000	  &	100,000			\\
$n$			&	17		&	40	&20   &	7			\\
$|\Pi_{QI}(R)|$		&	5,636		&	12,405 & 60	  &	24,630	 \\
$|\Sigma|$		&	24		&	 21	 & 18	  &	10		\\
\bottomrule
\end{tabular}
}
\end{center}
\end{table}

\starter{Datasets.}
We use three real data collections and one synthetic dataset.  Table~\ref{tb:dataset} gives the data characteristics, showing a range of data sizes w.r.t. the number of tuples ($|R|$), number of attributes ($n$), number of unique values in the QI attributes ($|\Pi_{QI}(R)|$), and the total number of defined diversity constraints ($|\Sigma|$). 

\eat{
\blue{\textbf{Forbes Richest.} This dataset contains Forbes Richest People lists from 2016: US Richest with 400 individuals (https://www.forbes.
com/forbes-400/list/) and World’s Richest with 526 individuals (https://www.forbes.com/billionaires/list/). They use two attributes as sensitive attributes: gender and country. Each of them has $2$ categories and $5$ categories, respectively. Although it is a small dataset, it is the benchmark dataset used in other fairness and diversity paper, which we will use it for baselien comparison. }
}
\dataset{Pantheon}~\cite{pantheon}. This dataset describes individuals based on the popularity of their biographical page in Wikipedia.  Attributes include \attr{name, sex, city, country, continent}. We select \attr{sex, city, country} and \attr{continent} as QI attributes, and define diversity constraints on  \attr{sex} and \attr{continent}, where the attribute domain is two and six, respectively.  \eat{, where the number of categories are 2 and 6. For each category, we consider three types of diversity constraints, so the total number of diversity constraints is 24.} We use this dataset to evaluate algorithm accuracy.  

\dataset{Census}~\cite{UCI}. The U.S. Census Bureau describes population data for 1970, 1980 and 1990.  We select \attr{sex, workclass, marital status, family relationship, race}, and \attr{native country} as QI attributes.  We define (single and multi-attribute) diversity constraints on the \attr{sex} and \attr{race} attribute domains with size two and five, respectively. We evaluate accuracy, runtime, and comparative performance with this dataset.\eat{, and the number of unique QI is 12,405 as Table \ref{tb:dataset} shows. We consider diversity constraints on sex and race, which has 2 and 5 categories separately. The total number of diversity constraints is 21.} \eat{I would suggest additions such as workclass, education, occupation and certainly race, these are diversity requirements normally.  The other statistic information of the dataset, includes defined diversity constraints, domain size for each attribute can be found on our external site \cite{dataset}.}

\dataset{German Credit}~\cite{UCI}. This dataset classifies persons as good or bad credit risk according to attributes such as  \attr{credit history, credit amount, sex, job, housing, marital status}, and stratified \attr{savings account} balances.  We select \attr{sex, job, housing, saving account} as QI attributes, and define diversity constraints on \attr{sex} and \attr{job} containing two and four values, respectively.  We comparatively evaluate against an existing $k$-anonymization baseline with this dataset.  



\dataset{Synthetic Population Data (Pop-Syn)}.
We use the Synner.io tool to generate realistic synthetic data by declaratively specifying the desirable distribution properties in the target attributes~\cite{MA19}.  We generate a synthetic dataset  describing population characteristics (\attr{age, education, race, gender, income, marital status, occupation}).  We select a subset of these attributes as target attributes, and vary their statistical distributions (uniform, Gaussian, Zipfian) to study the impact on \alg's accuracy. 



\starter{Diversity Constraints.} 
We implement different notions of diversity such as minimum frequency, average and proportional representation from the attribute domain.  We use the diversity definitions  presented by Stoyanovich et. al. that define three classes of diversity constraints as described below~\cite{stoyanovich2018online}.  We generate a set of satisfiable diversity constraints $\Sigma_i$  for each class, $i = \{1,2,3\}$, for each dataset.   

In the original definition, Stoyanovich et. al., define these diversity constraint classes w.r.t. the number of selected elements from a  set~\cite{stoyanovich2018online}.  In our setting, we consider an equivalent notion as the number $U$ of published (non-suppressed) tuples in $R'$.  To estimate $U$, recall the tuples in the QI attributes are suppressed to achieve the indistinguishability of a tuple among $(k-1)$ other tuples in a cluster group.  We can estimate $U$ by computing the cardinality of the QI attribute(s) domain, and subtracting this value from the size of $R$.  Let $\Pi_{QI}(R)$ represent the projection of relation $R$ on the QI attributes, i.e., the set of unique tuples w.r.t. the QI attributes.  These unique values will need to be suppressed among an average of $\frac{|R|}{k}$ groups to achieve $k$-anonymity.  Hence, we estimate $U = |R| - |\Pi_{QI}(R)|$ as the number of tuples that are published (unsuppressed) tuples in $R'$.  We now describe each class of diversity constraints.   Let $d = |dom(A)|$, i.e., the number of unique values in the target attribute(s) $A$ domain. The full set of diversity constraints, datasets and our code are available at~\cite{dataset}. 

\begin{itemize}[nolistsep, leftmargin=*]
    \item \textbf{Minimum:} Cover as many values in the attribute(s) domain as possible.  If $U > d$, set $\lambda_l = \lambda_r = 1$ for all $d$ (value) constraints. Then, compute $w = U - d$.  If $w > 0$, then assign these values to a random constraint $\sigma_j'$ by setting its $\lambda_r' = \lambda_r' + w$. Select $\sigma_j'$ randomly where \nit{freq}($a$) $\geq \lambda_r' + w$. 
    
    If $U < d$, set $\lambda_l = \lambda_r = 1$ to a random set of $U$ out of $d$ constraints, and set $\lambda_l = \lambda_r = 0$ to the remaining $d - U$ constraints.
    
    \item \textbf{Average:} Select equal numbers for each value in the attribute domain. 
    If $U \geq d$, set $\lambda_l = min(\lfloor U/d \rfloor, \nit{freq}(a))$, $\lambda_r = min(\lceil U/d \rceil, \nit{freq}(a))$, where $\nit{freq}(a)$ represents the frequency of value(s) $a$ in attribute(s) $A$ in $R$. Next, compute $w = \sum_{i=1}^d \lambda_{r_{i}}$. If $w < U$, then assign these values to a random $\sigma_j'$ by setting $\lambda_r' = \lambda_r' + w$. Select $\sigma_j'$ randomly where \nit{freq}($a$) $\geq \lambda_r' + w$.  If $U < d$, define as in \textbf{minimum} class. 
    \item \textbf{Proportion:} Select equal proportions for each value in $dom(A)$.  
    If $U \geq d$, set $\lambda_l = \lfloor U * \nit{freq}(a)/|R| \rfloor, \lambda_r = \lceil U * \nit{freq}(a) / |R| \rceil$.  If $U<d$, set constraints as in \textbf{minimum} class above.

\eat{
    \item \textbf{Relaxed average:}
    If $U>d$, set constraints as \textbf{average} above. Then set $\lambda_l = max( \lambda_l - t, 0), \lambda_r = min(\lambda_r + t,|A[a]|)$, where $t$ represents the tightness threshold which specified by user.
    
    If $U<d$, set constraints as in \textbf{minimum} constraint above.
    
    \item \textbf{Relaxed proportion:}
     If $U>d$, set constraints as \textbf{proportion} above. Then set $\lambda_l = max( \lambda_l - t, 0), \lambda_r = min(\lambda_r + t,|A[a]|)$.
    
    If $U<d$, set constraints as in \textbf{minimum} constraint above.
}
\end{itemize}

\starter{Comparative Baseline.} 
\noindent  
As far as we know, \alg is the first work to couple diversity and privacy-preserving anonymization.  The closest comparative baseline is the $k$-member anonymization algorithm takes a greedy, clustering-based approach to group similar records by minimizing the distance between values and between records~\cite{byun2007efficient}.  $k$-member aims to  minimize distortion among the values, and minimize the information loss in the anonymized relation.  Although $k$-member considers both generalization and suppression, we only apply suppression in our comparative evaluation.


\begin{figure*}[tb!]
	\captionsetup[subfloat]{justification=centering}
	\centering
		\subfloat[\small{Varying $\Sigma, k$ (Census)}]{\label{fig:acc_ksigma}
			{\includegraphics[width=0.24\linewidth]{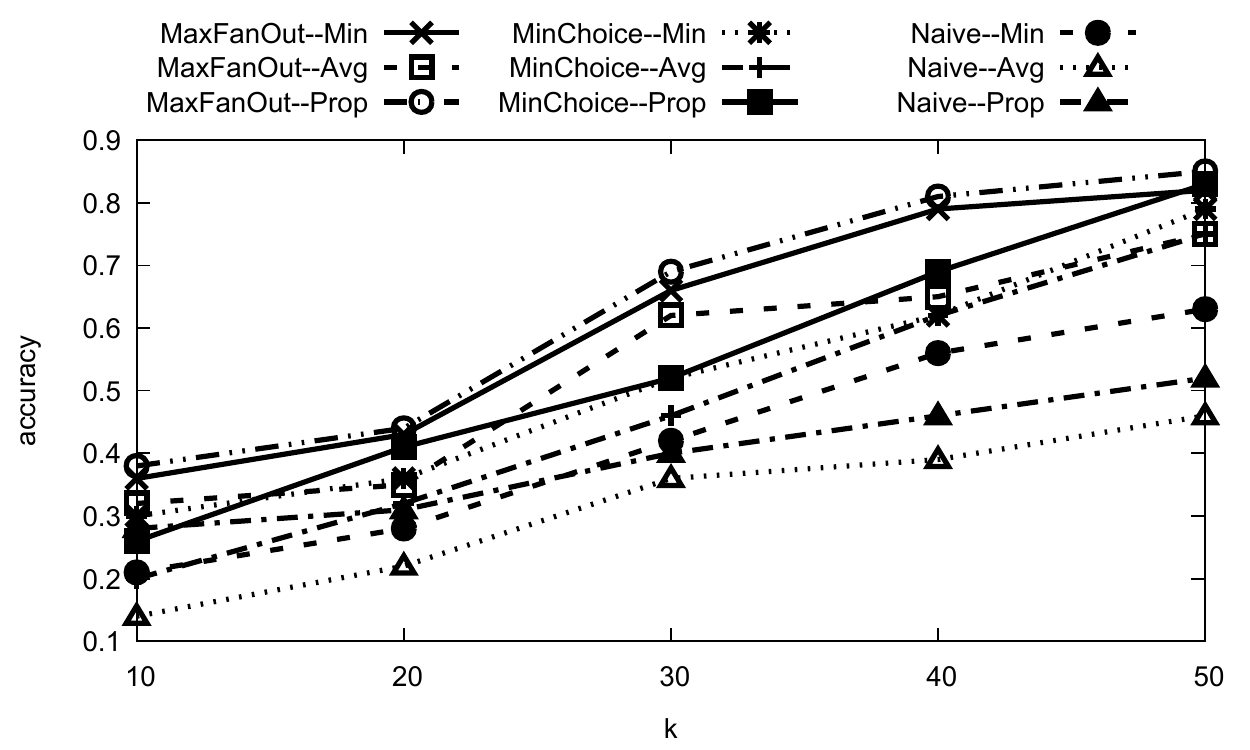}}}
		\hfill\subfloat[
		\small{Varying $\Sigma$, $\nit{cf}$ (Census)}]{\label{fig:acc_confsigma}
			{\includegraphics[width=0.24\linewidth]{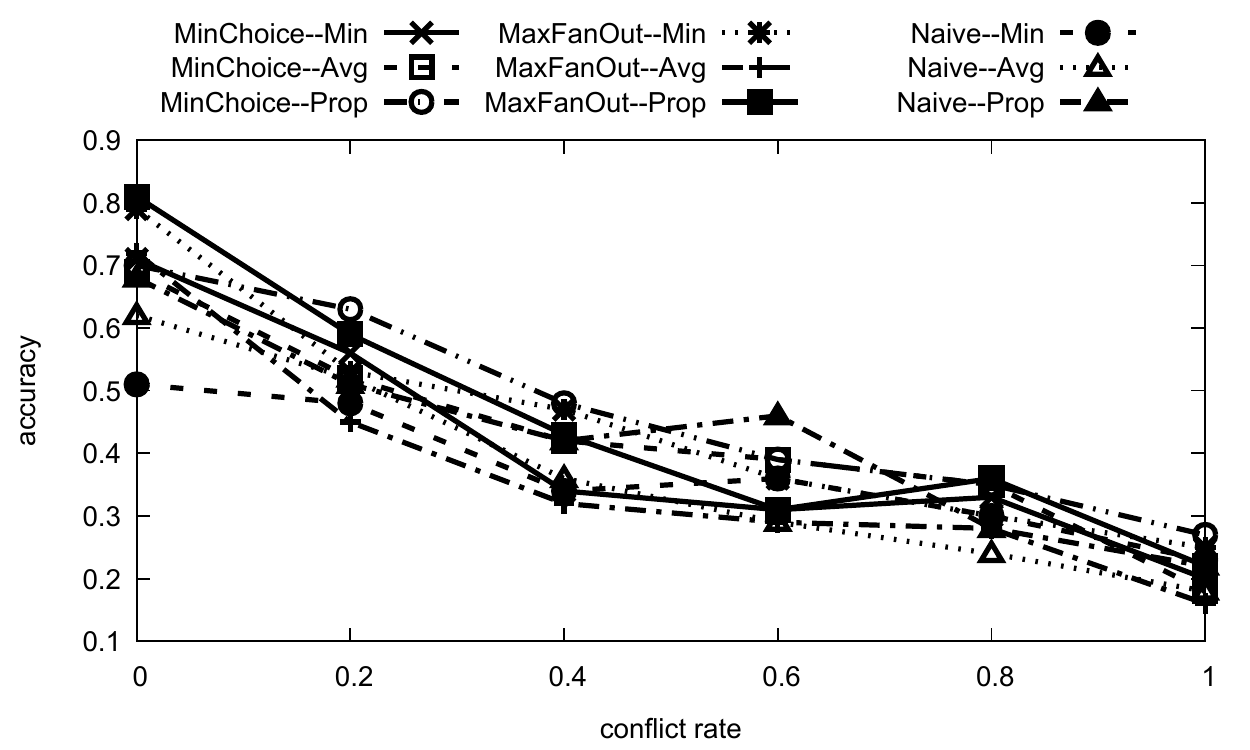}}}
		\hfill\subfloat[
		\small{Accuracy vs. $|\Sigma|$ (Pantheon)}]{\label{fig:acc_varySize1}
			{\includegraphics[width=0.24\linewidth]{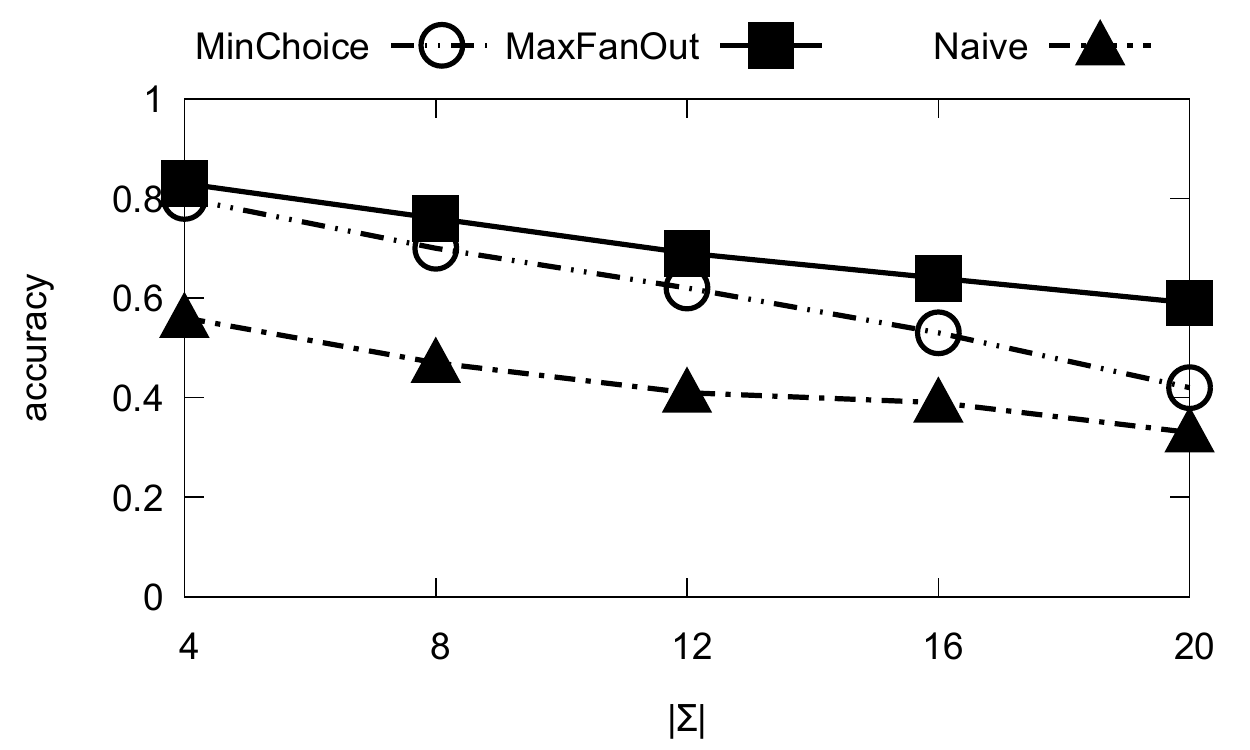}}}
		\hfill\subfloat[
		\small{Accuracy vs. $|\Sigma|$ (Census)}]{\label{fig:acc_varySize2}
			{\includegraphics[width=0.24\linewidth]{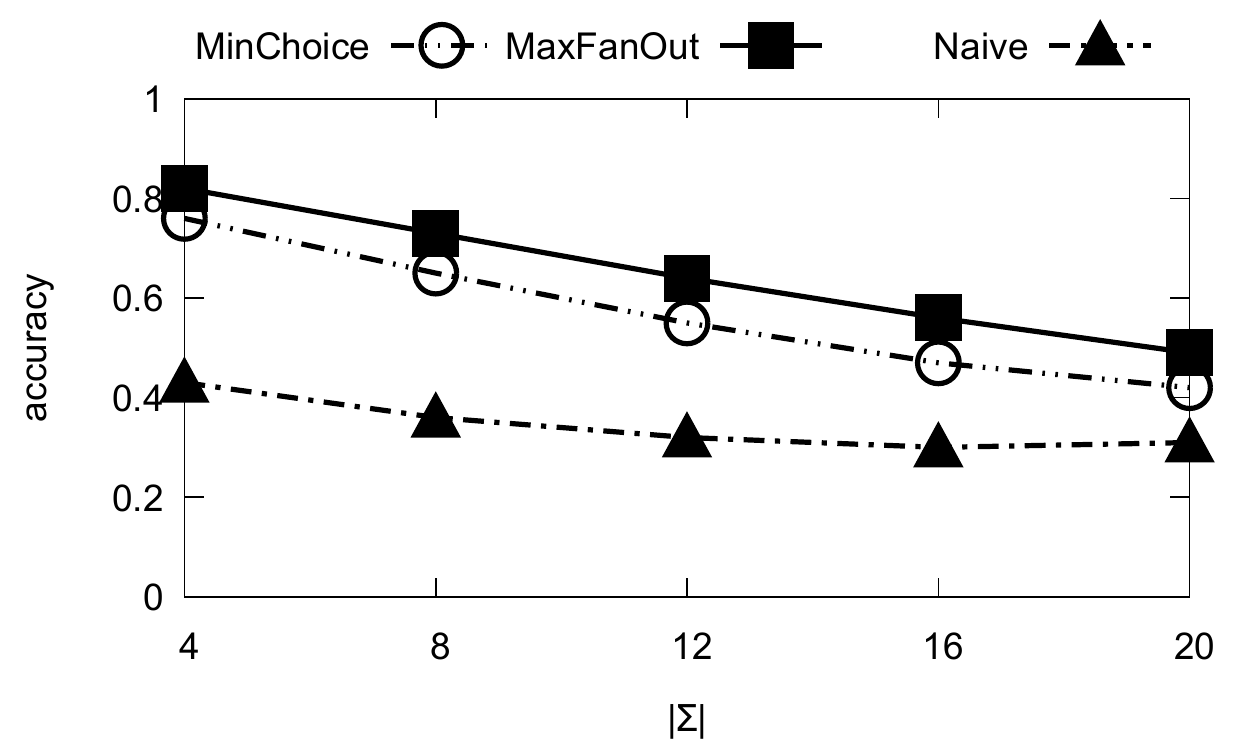}}}
		\hfill\subfloat[
		\small{Accuracy vs. $\nit{cf}$ (Pantheon)}]{\label{fig:acc_varyConf}
			{\includegraphics[width=0.24\linewidth]{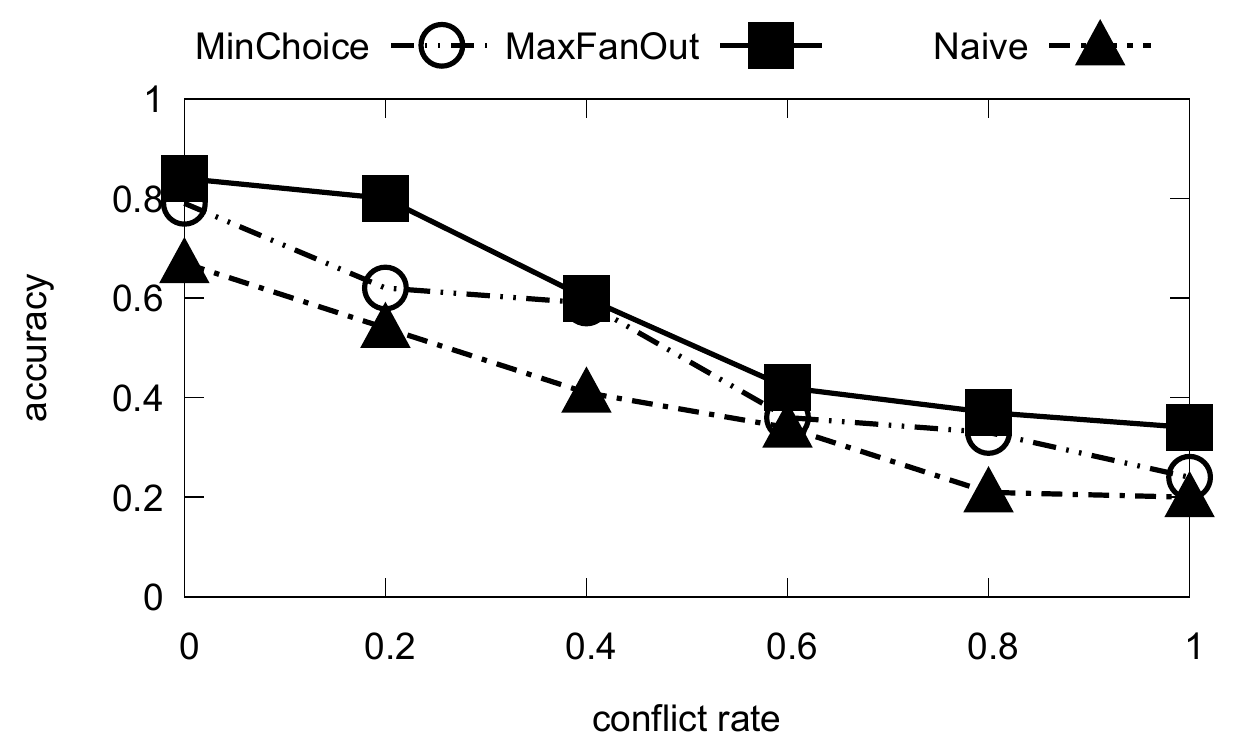}}}
		\hfill \subfloat[\small{Acc. vs. distrib. (Pop-Syn)}]{\label{fig:acc_varyData}
			{\includegraphics[width=0.24\linewidth]{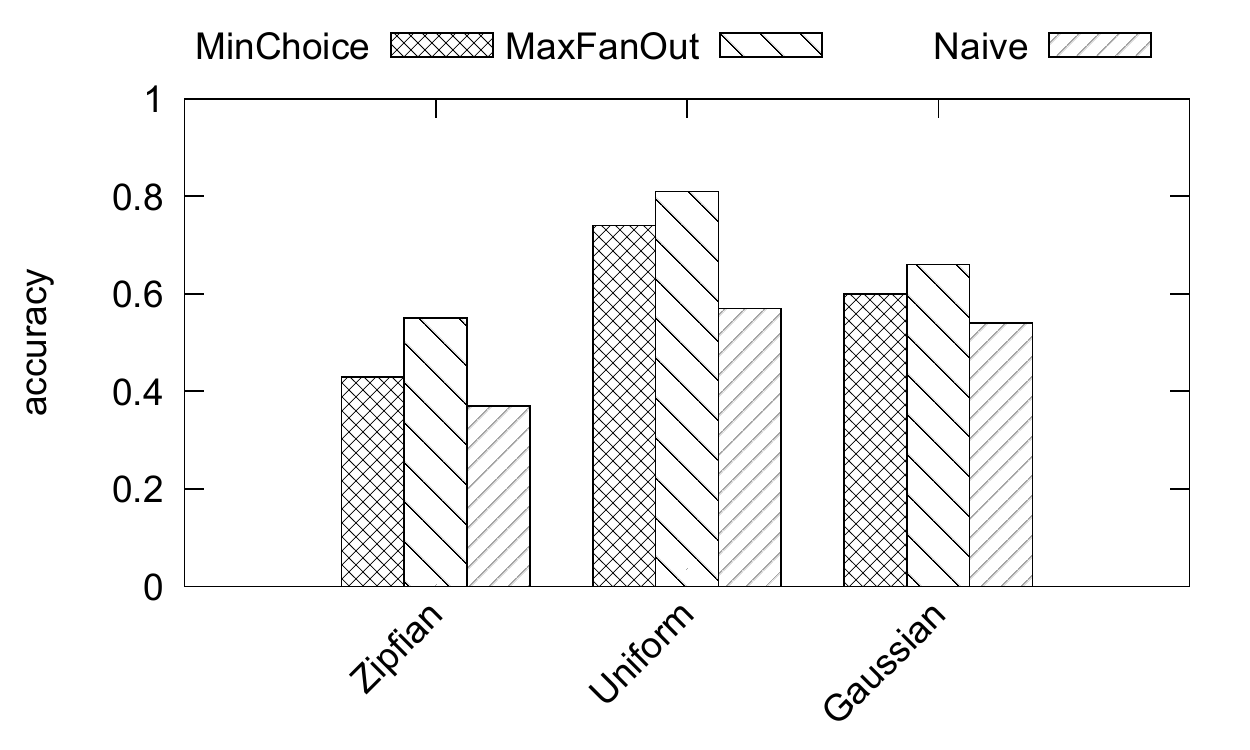}}}		
		\hfill\subfloat[
		\small{Runtime vs. $|\Sigma|$ (Census)}]{\label{fig:perf_varySize}
			{\includegraphics[width=0.24\linewidth]{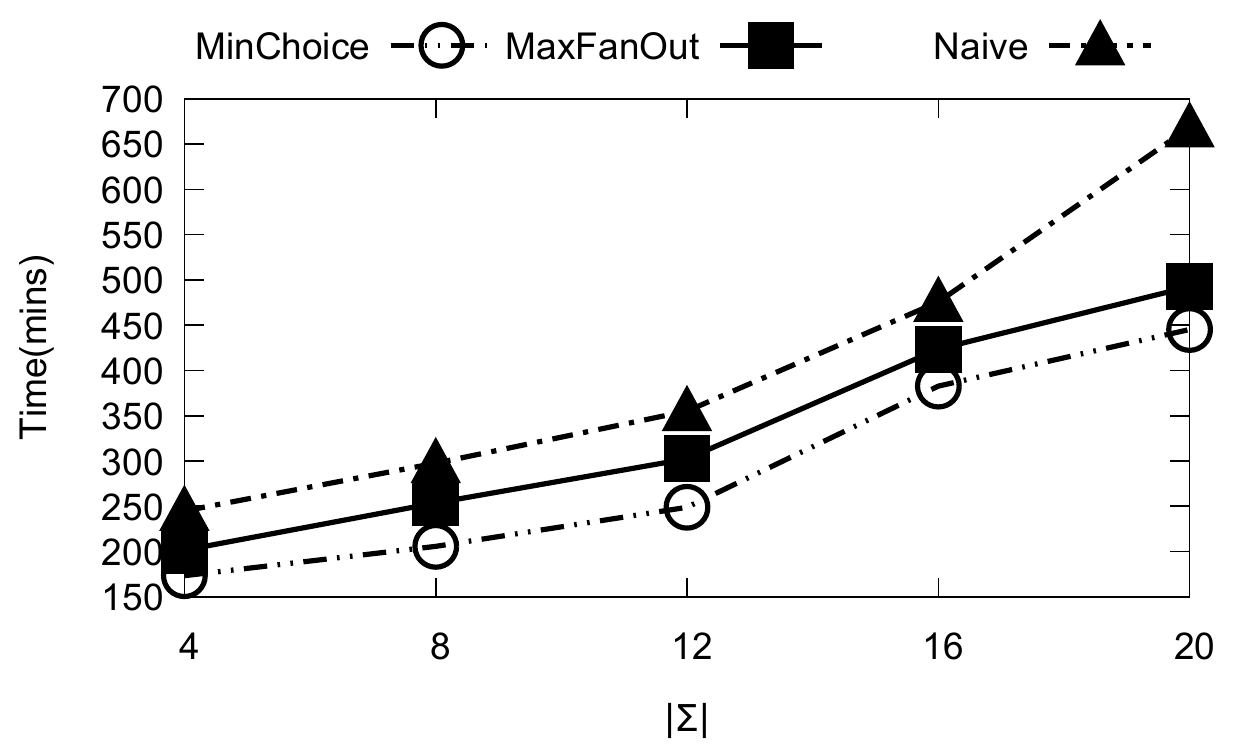}}}
		\hfill\subfloat[\small{Runtime vs. $\nit{cf}$ (Census)}]{\label{fig:perf_varyConf}
			{\includegraphics[width=0.24\linewidth]{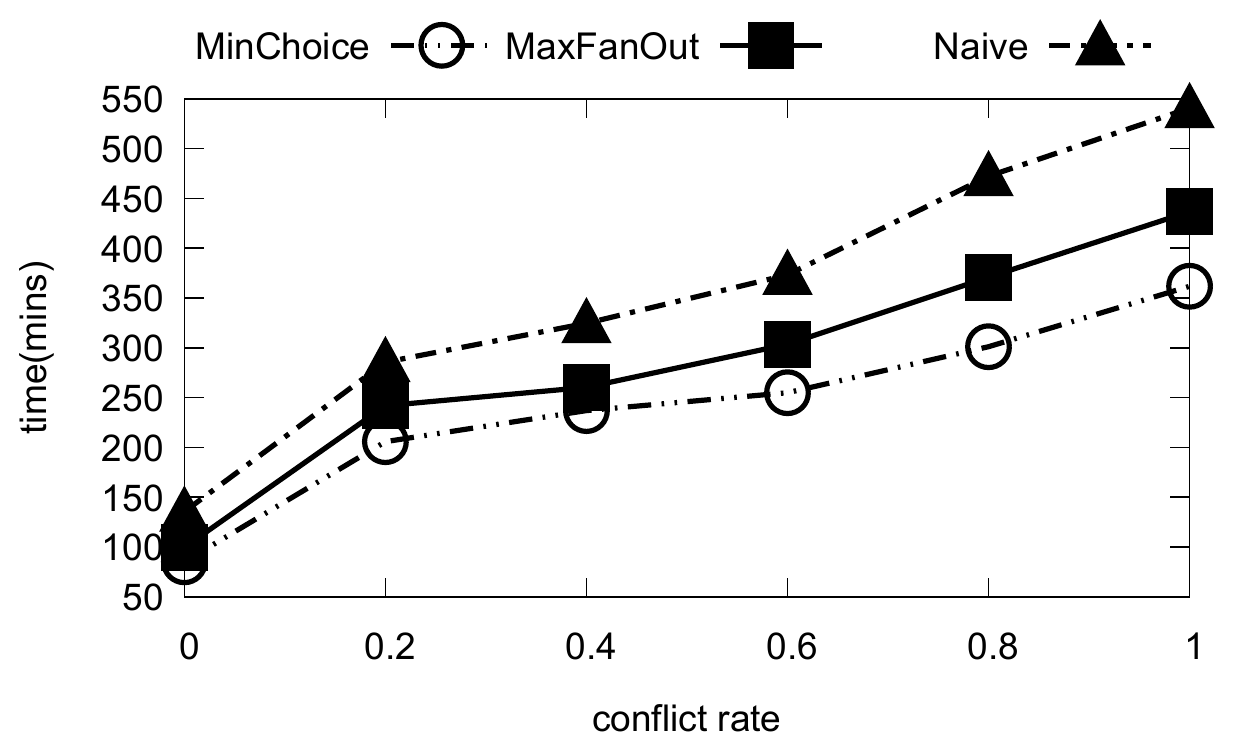}}}
	\vspace{-3ex}
	\caption{\alg effectiveness and efficiency.}\label{fig}
	\vspace{-2ex}
\end{figure*}

\begin{figure*}[tb!]
	\captionsetup[subfloat]{justification=centering}
	\centering
    	\subfloat[\small{ $\widehat{disc}$ vs. $k$ (Credit)}]{\label{fig:comp_kacc}
			{\includegraphics[width=0.24\linewidth]{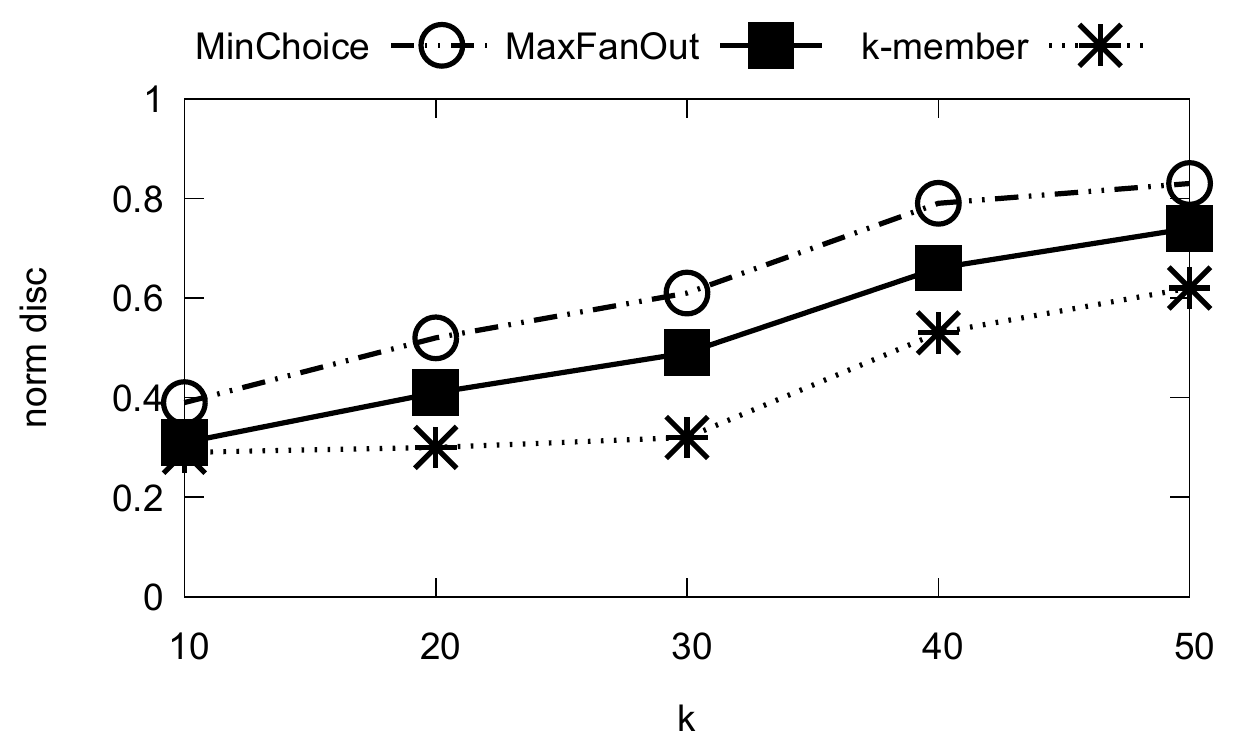}}}
		\hfill\subfloat[\small{Runtime vs. $k$  (Credit)}]{\label{fig:comp_kperf}
			{\includegraphics[width=0.24\linewidth]{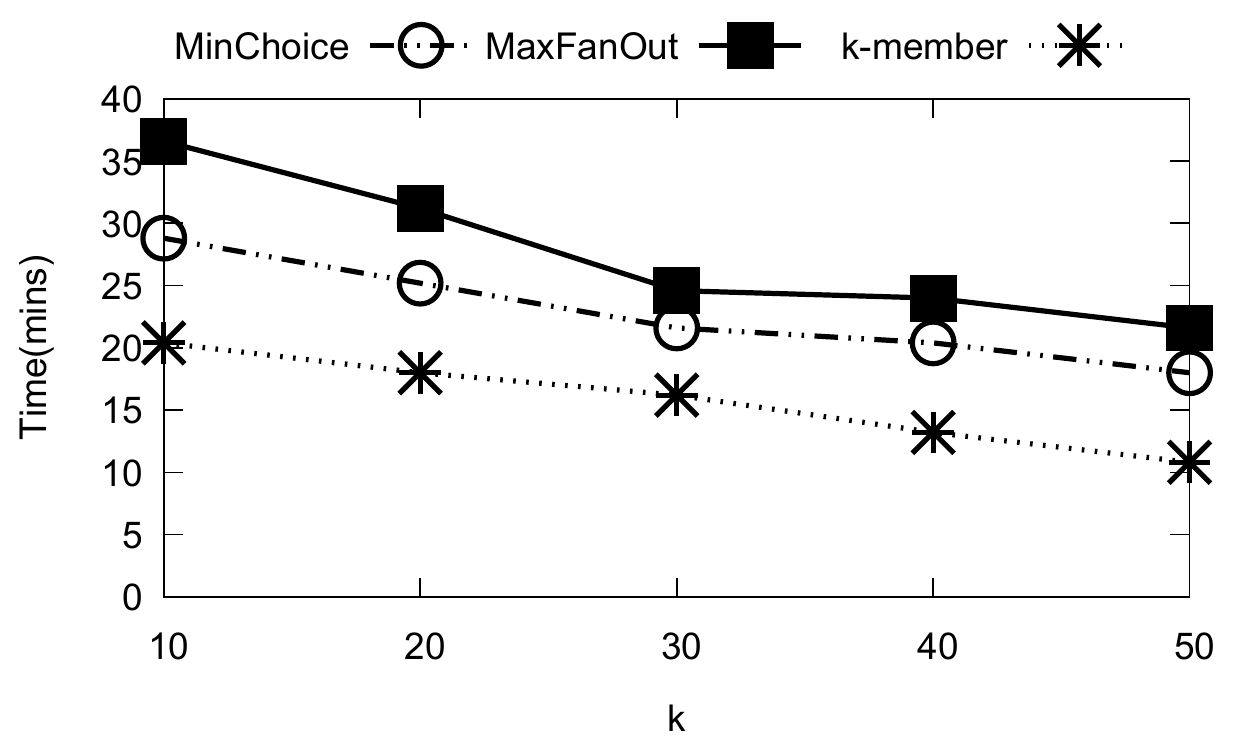}}}
		\hfill\subfloat[\small{ $\widehat{disc}$ vs. $|R|$ (Census) }]{\label{fig:comp_Racc}
			{\includegraphics[width=0.24\linewidth]{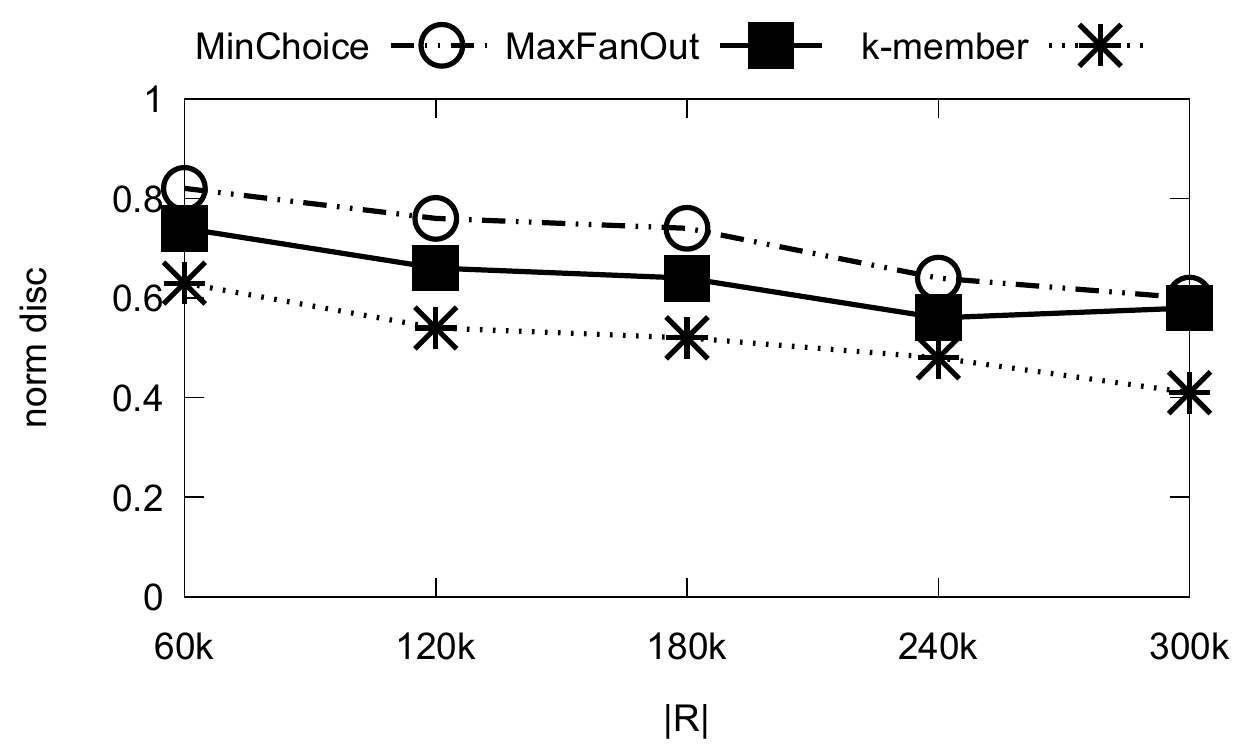}}}
		\hfill\subfloat[\small{Runtime vs. $|R|$ (Census)}]{\label{fig:comp_Rperf}
			{\includegraphics[width=0.24\linewidth]{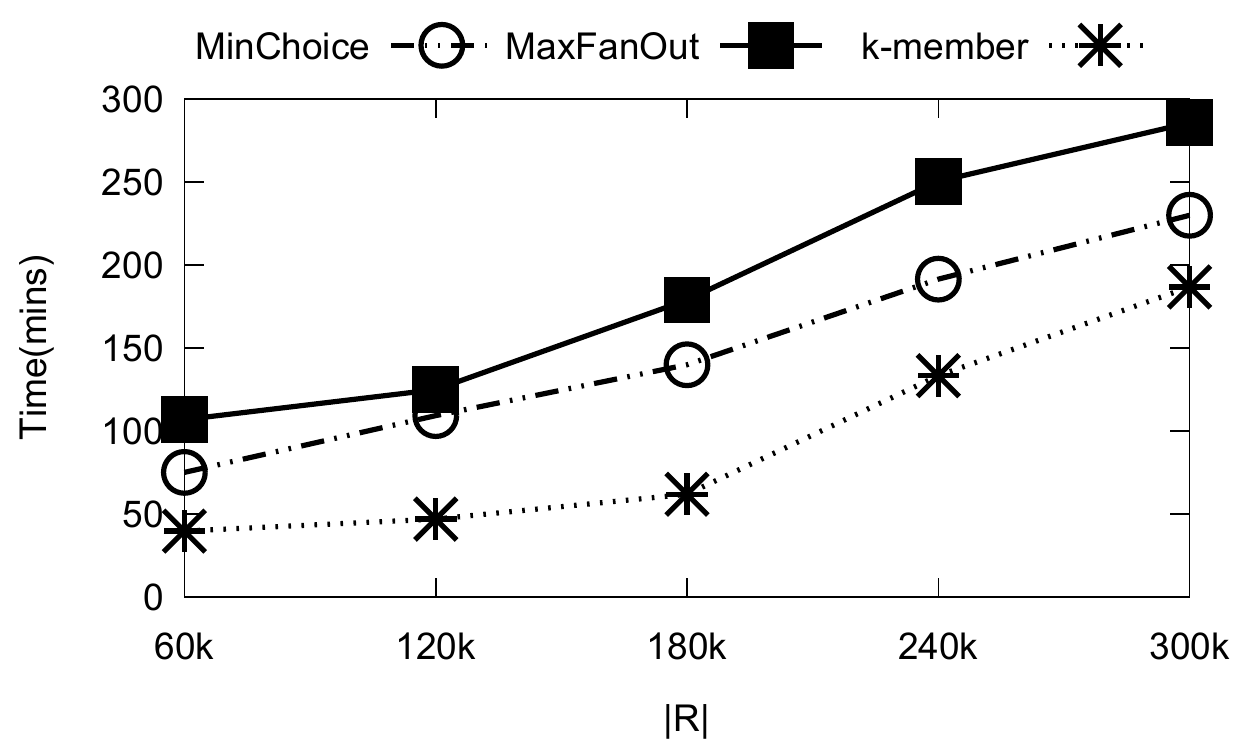}}}	
	\vspace{-2ex}
	\caption{Comparative evaluation. \eat{(1) Graph lines used for the same algorithm remain consistent across all graphs for all remaining figures 3c-3h, 4a-4d.  (2) Include all 3 versions of DIVA. (3) Increase all legend and x,y axes font size.} }\label{fig}
	\vspace{-2ex}
\end{figure*}



\begin{table}
\begin{small}
\begin{center}
\caption{Parameter values (defaults in bold) \label{tbl:defaults}}
\vspace{-4mm}
\begin{tabular}{ | l | l | l |}
          \hline
          \textbf{Symbol} & \textbf{Description} & \textbf{Values} \\
          \hline
          $|R|$  & \#tuples  & 60k, 120k, 180k, 240k, 300k \\
          \hline
          $|\Sigma|$ &  \#constraints & 4, \textbf{8}, 12, 16, 20 \\
          \hline
           $\nit{cf}(\Sigma)$ & conflict rate & 0, \textbf{0.2}, 0.4, 0.6, 0.8, 1 \\ 
          \hline
          $k$  & minimum cluster size & 10, \textbf{20}, 30, 40, 50 \\
          \hline
\end{tabular}
\end{center}
\end{small}
\end{table}

\subsection{Metrics and Parameters}

\starter{Metrics.} We compute the average runtime over five executions. To quantify accuracy, we use an intuitive measure to model  desirable anonymizations that minimize a cost function.  Existing anonymization algorithms use cost functions that minimize information loss from suppression~\cite{bng-fung\ignore{,BA05,LDR06}}.  The resulting anonymized relation $R'$ can be considered as imposing a penalty on each tuple that reflects its information loss due to suppression.  The \emph{discernibility metric}, $disc(R', k)$, quantifies the differentation between tuples for a given $k$ value, by assigning a penalty to each tuple based on the number of tuples that are indistinguishable from it in $R'$~\cite{BA05}.  If an  unsuppressed tuple lies in a cluster of size $j$, it is assigned a penalty of $j$.  If a tuple is suppressed, it is assigned a penalty of $|R'|$ since the tuple cannot be differentiated from other tuples in $R'$~\cite{BA05}. We define the normalized discernibility score as   $\widehat{disc}(R',k) = \frac{disc(R', k)}{|R'^2|}$.  To quantify accuracy, we compare $disc(R', k)$ for $R'$ computed by \alg against $disc(R'', k)$ for the best $R''$ computed by sampling among all the possible clusters and selecting the best clustering.  We compute accuracy as the ratio of the normalized discernibility scores $\frac{\widehat{disc}(R'',k)}{\widehat{disc}(R',k)}$.  In our comparative evaluation, we measure accuracy using 
 $\widehat{disc}(R',k)$ to quantify the penalty to enforce diversity in $R'$.


\starter{Parameters.}  Unless otherwise stated, Table~\ref{tbl:defaults} shows the range of parameter values we use, with default values in bold.  We measure the \emph{conflict rate} between the diversity constraints by measuring the number of overlapping relevant tuples between a pair of diversity constraints.  We use Jaccard similarity to quantify the similarity between two sets, computed as the size of the intersection divided by the size of the union of the sets.  Similarly, we define the conflict rate $\nit{cf}(\sigma_i, \sigma_j)$ = $\frac{|I_{\sigma_i} \cap I_{\sigma_j}|}{|I_{\sigma_i} \cup I_{\sigma_j}|}$ between constraints $\sigma_i, \sigma_j$, and $I_{\sigma_i}$ refers to the relevant tuples of $\sigma_i$.  For all $\sigma_i \in \Sigma$, we compute $\nit{cf}(\Sigma) = \frac{\Sigma_{i=1}^{|\Sigma|}\nit{cf}(\sigma_i, \sigma_{i+1})}{{|\Sigma| \choose 2}}$, i.e., the average of all conflict scores for every pair of diversity constraints.  Values of $\nit{cf}(\Sigma)$ range from [0, 1], where 0 indicates no overlapping relevant tuples, and 1 indicates full overlap (exact similarity) of the relevant tuples among the constraints.


\subsection{Accuracy}
We evaluate accuracy using three classes of constraints, and then vary $|\Sigma|$, $cf$, and the data distribution in the target attribute values.

\experiment{Exp-1: Vary $\Sigma$ and $k$.} Figure~\ref{fig:acc_ksigma} gives the \alg accuracy for the three variations of \alg as we vary $k$ using the Census dataset, across the three diversity constraint classes.  Accuracy increases for larger $k$ values as more values are suppressed to achieve anonymization. As expected, \stratNaive leads to the lowest accuracy due to its random selections.  \stratTwo outperforms \stratOne by an average +9\%, since by ordering clusterings in ascending order according to the number of overlapping tuples, we select clusterings that satisfy a maximal number of dependent constraints.  In contrast, \stratOne does not consider this constraint interaction.  \eat{reason for minChoice better than maxFanout? Yu to add text. \blue{Because \stratTwo begins with constraint that overlaps with the highest number of interactions with other constraints, which leads to a larger solution space than \stratOne. Therefore, \stratTwo is more possible to find a better solution than \stratOne which starts with fewest candidate clusterings to limit the size of solution space.} \fei{opposite the pt I try to make, how does this impact accuracy?}} The proportion class of constraints achieves the best tradeoff between accuracy and adapting to the relative frequency of values in the data.  Although the minimum class of constraints achieves higher accuracy in some cases (given the minimal lower bound values), this can lead to tokenization in $R'$.  

\eat{Three variations of \alg perform better on Minimum diversity constraints than the other two diversity classes, this is because the minimal type of constraint only requires minimal coverage for each category, which is easy to be satisfied. We also observe that as $k$ increases the accuracy of three \alg variations increase as well. This is because when $k$ is small, there are few number of tuples that are be suppressed, and the normalized discernibility score $\widehat{disc}(R',k)$ is small as well.  \fei{I can see this being true for minimality constraints, but not for the other types.} As $k$ increases, we need to suppress more tuples to satisfy the anonymity requirement, which makes the anynonymized relation $R'$ close to the best possible $R''$ w.r.t the number of suppressed tuples.} 


\experiment{Exp-2: Vary $\Sigma$ and Conflict Rate.} Figure~\ref{fig:acc_confsigma} shows \alg accuracy as we vary the conflict rate ($\nit{cf}$) across the three constraint classes. Accuracy declines for increasing $\nit{cf}$ as it is more difficult to find a clustering.  Again, \stratNaive achieves the lowest accuracy, whereas \stratTwo performs best by first selecting clusterings that satisfy neighboring vertices (constraints).  \eat{Among the constraint classes, we find that the minimum class of constraints (which are easier to satisfy), the average, and proportion class each achieve an average \tbf\%,  \tbf\%, \tbf\%  \blue{(with Naive 53\%, 42\%, 50\%, without Naive 61\%, 49\%, 57\% )}\fei{update for $cf = 0.2$ with and without Naive.} accuracy, respectively, at $cf = 0.2$, over all algorithm variations.}  The proportion class of constraints capture the relative distribution in the attribute domain (with less sensitivity than average), and avoids tokenization (a drawback of minimum constraints). Henceforth, we run subsequent experiments using the proportion class constraints.

\experiment{Exp-3: Vary $|\Sigma|$.} Figure~\ref{fig:acc_varySize1} and Figure~\ref{fig:acc_varySize2} show  the \alg accuracy as we vary the number of (proportion) constraints $|\Sigma|$ using the Pantheon and Census dataset, respectively.  \stratTwo outperforms \stratNaive and \stratOne by +27\% and +9\%, respectively, (Pantheon), and +30\% and +7\% (Census).  \eat{This difference can be explained by \blue{different search strategies of these three approaches. \stratTwo always begins with a larger search space than the other two, which can potentially find better solutions.}}  As $|\Sigma|$ increases, we see accuracy decline but at a relatively slow linear rate.  As a new constraint $\sigma \not \in \Sigma$ is added, we observe new relevant tuples w.r.t. $\sigma$ join existing clusters of relevant tuples from $\Sigma$ leading to a smaller decline in accuracy. This occurs with multi-attribute constraints that share target attributes with single attribute constraints.  The alignment of QI and target attribute values between new and existing tuples influence the accuracy rate of decline.

\experiment{Exp-4: Vary Conflict Rate.} Figure~\ref{fig:acc_varyConf} shows the \alg accuracy as we vary the conflict rate ($\nit{cf}$).  As expected, accuracy declines for increasing $\nit{cf}$, with \stratTwo and \stratOne outperforming \stratNaive by +17\% and +9\%, respectively. \stratTwo shows improved accuracy over \stratOne since targeting constraints with a high number of interactions (with other constraints) first allows it to eliminate unsatisfying clusterings sooner, while also satisfying dependent diversity constraints.


\experiment{Exp-5: Vary Data Distribution.} \eat{We study the impact of varying data distributions on accuracy.}  We  generate target attribute values according to the Zipfian, uniform, and Gaussian distributions in the Pop-Syn dataset with $|R| = 100k$ and $|\Sigma| = 8$. Figure~\ref{fig:acc_varyData} shows that \stratTwo performs best across all distributions by 8\% and 17\% over \stratOne and \stratNaive, respectively.  The target uniform distribution performs best as domain values are spread evenly across the tuples, avoiding contention among a small set of tuples.  This conflict occurs more often in the Zipfian case than the Gaussian, leading to lower accuracy. 
\subsection{Performance} 
\label{sec:exp_perf}



\experiment{Exp-6: Scale $|\Sigma|$.} Figure~\ref{fig:perf_varySize} shows the \alg runtime as we vary the number of constraints over the Census dataset.  As expected, \stratNaive shows exponential growth for increasing $|\Sigma|$ since we can assign $O(|R|)$ different clusterings to each constraint.  Our selection strategies to restrict clusterings and perform early pruning in \stratOne and \stratTwo show linear scale-up with a 29\% and 18\%, respectively, reduction in runtime over the naive version. 


\experiment{Exp-7: Vary Conflict Rate.} Figure~\ref{fig:perf_varyConf} shows runtimes as we vary the conflict rate.  \stratOne outperforms \stratTwo and \stratNaive by 16\% and 23\%, respectively.  We observe that when conflicts occurs among a set of tuples, leaving residual tuples that are unique and the only ones that can satisfy a constraint, e.g., vertex $v_2$ ($\sigma_2$) in Figure~\ref{fig:graph}, \stratOne performs well.  By selecting these special constraints first (with fewer clustering choices), we reduce the number of clusterings to evaluate.

\subsection{Overhead of Diversity Constraints}

\experiment{Exp-8: Vary $k$.}  Figure~\ref{fig:comp_kacc} and Figure~\ref{fig:comp_kperf} show the comparative discernibility scores and runtimes between \alg and $k$-member~\cite{byun2007efficient}.  \stratOne and \stratTwo incur an average 32\% and 44\% higher runtime, respectively, than $k$-member, reflecting the cost of computing a \emph{diverse} data instance.  As $k$ increases, we expect more tuples to be suppressed leading to higher penalty costs, and higher $\widehat{disc}(R',k)$ scores.  For \stratTwo and \stratOne, a 10\% reduction in $\widehat{disc}(R',k)$ costs 13m and 9m, respectively, whereas for $k$-member, the cost is 4m.  We believe that the overhead and trade-off are still acceptable in practice since constraint validation and anonymization is often done offline.  As next steps, we are exploring techniques to reduce the overhead via parallel processing of the \nit{Coloring} routine on subgraphs of $G$.


\experiment{Exp-9: Vary $|R|$.}  Figure~\ref{fig:comp_Racc} shows that as $|R|$ increases,  $\widehat{disc}(R',k)$ scores slightly improve as QI and target attribute values from the new tuples align with existing tuples, and do not incur additional suppression (penalty).  In contrast, when new attribute values are suppressed to satisfy diversity constraints (at $|R| = 240K$), we incur increased penalty costs. \eat{(1) the upper $\lambda_r$ bounds being violated causing more unsatisfied constraints; and (2) an increased number of suppressed values are needed to achieve alignment along QI attributes.}  Figure~\ref{fig:comp_Rperf} shows that \alg runtimes increase linearly w.r.t $|R|$ with an average overhead of 36\% over the baseline, as new tuples and clusterings need to be evaluated.

%% file: related-work.tex
\section{Related Work} \label{sec:rw}

\starter{Privacy Preserving Data Publishing.}
Extensions of $k$-anonymity include {\em $l$-diversity},  {\em $t$-closeness}, {\em (X,Y)-privacy}, and  {\em (X,Y)-anonymity} with tighter privacy guarantees~\cite{bng-fung}.  \alg is extensible to re-define the clustering criteria according to these privacy semantics. {\em Differential privacy (DP)} provides a higher level of protection for individuals where the existence (or not) of a single record should not impact the outcome of any statistical analysis~\cite{dwork}. As next steps, we intend to study similar decision problems, and quantify the randomization to satisfy both DP and a set of diversity constraints. Cuenca et. al, study PPDP in linked data by formalizing the anonymization problem and its complexity for RDF graphs~\cite{GK16}. Hay et. al., present a data publishing algorithm that guarantee anonymity over social network data~\cite{hay}. \eat{e.g. social network data about individuals, their relationships, and shared activity.} In generalization, data values are replaced with less specific, but semantically consistent values according to a generalization hierarchy~\cite{bng-fung}.  While \alg currently considers suppression (a special case of generalization), we are exploring distance metrics to include generalization in \alg.

\starter{Fairness and Diversity.}
Achieving fair and equal treatment of groups and individuals is difficult in data-driven decision making~\cite{BS16}.  Despite a strong need for algorithmic fairness and data diversity, such principles are rarely applied in practice~\cite{Sween13}.  Data sharing of private data has been studied along two primary lines. First, causality reasoning aims to recognize discrimination to achieve algorithmic transparency and fairness.  Recent techniques have proposed influence measures to identify correlated attributes~\cite{ASZ16}, statistical reasoning about discrimination~\cite{NS18}, and reasoning between causality and fairness to generate bias-free, differentially private synthetic data~\cite{YRK19}.  Secondly, recent work have studied variants of DP to release synthetic data with similar statistical properties to the input data~\cite{BS17}, publishing differentially private histograms~\cite{XZX13}, and studying the impact of differentially private algorithms on equitable resource allocation, especially for strict privacy-loss budgets~\cite{PMK20}.   Our work is complementary to these efforts, with a different goal; to publish diverse and anonymized versions of the original data with minimal information loss for applications where statistical summaries, synthetic data, and aggregate queries are inadequate.  Recent work by Stoyanovich et. al., study diversity in the set selection problem and introduce diversity constraints to guarantee representation for each category in the selected set~\cite{YS17,stoyanovich2018online}.  We build upon this work, and are the first to formalize diversity constraints and study their foundations.  We  propose algorithms to couple diversity with data anonymization, a problem not considered in existing work.

\starter{Diverse Clustering.} 
\eat{Clustering involves partitioning a set of points into disjoint clusters that satisfy either cluster-based or instance level constraints such as enforcing a minimum cluster size or the total number of clusters.} Incorporating diversity into  clustering has been limited to producing more diverse results.  Nguyen et. al., start with an initial clustering and then generate additional clusterings that minimize error from the initial set~\cite{CENS06}.  Phillips et. al., argue that there is limited success by being too reliant on the initial clustering, and propose a sampling approach to select a diverse, large sample of non-redundant clusters while maximizing a quality metric~\cite{PRV11}.  The only work we are aware of that combines clustering with anonymization is by Li et. al., that study a 2-approximation algorithm for $l$-diversity, an extension of $k$-anonymity, where each cluster is of size at least $l$, and each point is a different color (i.e., sensitive value)~\cite{LYZ10}.   However, while our work shares a similar spirit, Li et. al., show that a solution may not be possible depending on the color distribution, and record deletion may be necessary.  \alg does not consider tuple deletion, and we use graph coloring to model tuple overlap between constraints, focusing instead on a declarative specification of diversity that is realizable in practice.

%% file: conclusion.tex
\section{Conclusion} \label{sec:conclusion}

We introduce \alg, a \uline{DIV}ersity-driven \uline{A}nonymization algorithm that computes a privatized data instance guaranteed to satisfy a set of diversity constraints.  We studied the foundations of diversity constraints, and presented a sound and complete axiomatization. \eat{uses a minimal amount of suppression to safeguard sensitive values, while ensuring the privatized instance satisfies a set of diversity constraints.}  We showed that the $(k,\Sigma)$-anonymization decision problem is in PTIME, presented a clustering-based algorithm, and proposed optimizations to improve performance.  \eat{After studying the algorithmic and theoretical foundations, our natural next steps include evaluating the algorithm efficiency, and effectiveness using real datasets.} Our evaluation showed the performance benefits of the optimizations, and the overhead of enforcing diversity constraints over the baseline.  As future work, we intend to study \eat{do we want to keep this or talk about generalization and distributed processing?} more expressive statistical-based diversity constraints, and privacy extensions beyond $k$-anonymity. We are also investigating  a distributed version of the coloring algorithm in \nit{DiverseClustering} for improved scalability.